\documentclass[11pt]{article}
\pdfoutput=1

\usepackage{jheppub}
\usepackage{comment}
\usepackage{times}
\usepackage{latexsym}
\usepackage{graphicx, graphics, amsmath, amssymb, slashed, bbm,bm,amsthm, array}
\usepackage{subfigure}
\usepackage{listings} 
\usepackage{trimclip}
\usepackage[dvipsnames]{xcolor}
\usepackage{hyperref}
\usepackage{xspace}
\usepackage{rotating}
\usepackage{afterpage}
\usepackage{multirow} 
\usepackage{lineno}
\usepackage{arydshln}
 
\newcommand{\nc}{\newcommand}

\nc{\mc}{\mathcal}
\nc{\er}[1]{(\ref{eq:#1})}
\nc{\onehalf}{\frac{1}{2}} 
\nc{\partialbar}{\bar{\partial}}
\nc{\psit}{\widetilde{\psi}}
\nc{\Tr}{\mbox{Tr}}
\nc{\hc}{\mbox{H.c.}}
\nc{\ev}{\ensuremath{\text{e\kern -0.1em V}}\xspace}
\nc{\mev}{\ensuremath{\text{Me\kern -0.1em V}}\xspace}
\nc{\gev}{\ensuremath{\text{Ge\kern -0.1em V}}\xspace}
\nc{\kev}{\ensuremath{\text{ke\kern -0.1em V}}\xspace}
\nc{\tev}{\ensuremath{\text{Te\kern -0.1em V}}\xspace}
\nc{\pev}{\ensuremath{\text{Pe\kern -0.1em V}}\xspace}
\nc{\eev}{\ensuremath{\text{Ee\kern -0.1em V}}\xspace}
\nc{\red}[1]{\textcolor{red}{#1}}

\def\chii0{\chi_i^0}
\def\chij0{\chi_j^0}

\newcommand{\HT}{\ensuremath{H_{\text{T}}}\xspace}
\newcommand{\met}{\ensuremath{E_{\text{T}}^{\text{miss}}}\xspace}
\newcommand{\pt}{\ensuremath{p_{\text{T}}}\xspace}
\newcommand{\gsim}{\lower.7ex\hbox{$\;\stackrel{\textstyle>}{\sim}\;$}}
\newcommand{\lsim}{\lower.7ex\hbox{$\;\stackrel{\textstyle<}{\sim}\;$}}
\nc{\ttbar}{t\bar t}

\newcommand{\cref}[1]{Chapter~\ref{c.#1}}

\hypersetup{
    colorlinks=true,
    linkcolor=blue,
    filecolor=magenta,
    urlcolor=cyan,
}

\def\DC#1{}

\title{Review of opportunities for new long-lived particle triggers in Run 3 of the Large Hadron Collider}

\author[]{\phantom{a} \\Produced for the LPCC Long-Lived Particles Working Group. 
\\
\phantom{a} \\
Editors:\\}

\author[5]{\phantom{a}\\Juliette Alimena,}
\author[4]{James Beacham,}
\author[3]{Freya Blekman,}
\author[11]{Adrián Casais Vidal,}
\author[11]{Xabier Cid Vidal,}
\author[31]{Matthew Citron,}
\author[36]{David Curtin,}
\author[5]{Albert De Roeck,}
\author[24]{Nishita Desai,}
\author[8]{Karri Folan Di Petrillo,}
\author[21]{Yuri Gershtein,}
\author[10,27,5]{Louis Henry,}
\author[35]{Tova Holmes,} 
\author[10]{Brij Jashal,}
\author[33]{Philip James Ilten,}
\author[6]{Sascha Mehlhase,}
\author[5]{Javier Montejo Berlingen,}
\author[10]{Arantza Oyanguren,}
\author[28]{Giovanni Punzi,}
\author[26]{Murilo Santana Rangel,}
\author[9]{Federico Leo Redi,}
\author[13]{Lorenzo Sestini,}
\author[10]{Emma Torro,}
\author[5]{Carlos Vázquez Sierra,}
\author[37]{Maarten van Veghel,}
\author[17]{Mike Williams,}
\author[10]{Jos\'e Zurita.}

\author[]{\\ \phantom{a} \\ Contributing and supporting authors:\\}

\author[20]{\phantom{a}\\Darin Acosta,}
\author[16]{Christina Agapopoulou,}
\author[23]{Sara Alderweireldt,}
\author[8]{Artur Apresyan,}
\author[16]{Aurélien Bailly-Reyre,}
\author[5]{Carla Marín Benito,} 
\author[13]{Alessandro Bertolin,} 
\author[7,16]{Lukas Calefice,} 
\author[29]{Daniel Hugo Cámpora Pérez,} 
\author[12]{Andrea Coccaro,}
\author[20]{Sven Dildick,}
\author[22]{Caterina Doglioni,}
\author[23]{Sinead Farrington,}
\author[23]{Jack Gargan,}
\author[15]{Stefano Giagu,}
\author[25]{Jason Gilmore,}
\author[16]{Vladimir Gligorov,} 
\author[23]{Giulia Gonella,}
\author[23]{Guillermo Hamity,}
\author[16]{Arthur Hennequin,} 
\author[25]{Tao Huang,}
\author[8]{Sergo Jindariani,}
\author[17]{Daniel Johnson,} 
\author[30]{Georgia Karapostoli,}
\author[23]{Aaron Kilgallon,}
\author[19]{Gillian Kopp,}
\author[8]{Martin Kwok,}
\author[30]{Owen Long,}
\author[34]{Jia Fu Low,}
\author[39]{Steven Lowette,}
\author[1]{Jingyu Luo,}
\author[34]{Nik Menendez,}
\author[11]{Titus Mombächer,} 
\author[20]{Paul Padley,}
\author[23]{Victoria Parrish,}
\author[8]{Cristián Peña,}
\author[13]{Lorenzo Pica,} 
\author[38]{Vladimir Rekovic,}
\author[34]{Suzanne Rosenzweig,}
\author[25]{Alexei Safonov,}
\author[5]{Jakob Salfeld-Nebgen,}
\author[2]{María Spiropúlu,}
\author[32]{Giulia Tuci,}  
\author[19]{Chris Tully,}
\author[18]{Andrii Usachov,} 
\author[17]{David Vannerom,}
\author[23]{Akanksha Vishwakarma,}
\author[2]{Christina Wang,}
\author[2]{Si Xie,}
\author[34]{Efe Yiğitbaşı,}
\author[23]{Estifa’a Zaid.}

\affiliation[1]{Brown University, Providence, RI, United States}
\affiliation[2]{California Institute of Technology, Pasadena, CA, United States}
\affiliation[3]{Deutsches Elektronen-Synchrotron (DESY), Hamburg, Germany}
\affiliation[4]{Duke University, Durham, NC, United States}
\affiliation[5]{European Organization for Nuclear Research (CERN), Geneva, Switzerland}
\affiliation[6]{Fakultät für Physik, Ludwig-Maximilians-Universität München, München, Germany}
\affiliation[7]{Fakultät Physik, Technische Universität Dortmund, Dortmund, Germany}
\affiliation[8]{Fermi National Accelerator Laboratory, Batavia, IL, United States}
\affiliation[9]{Institute of Physics, Ecole Polytechnique Fédérale de Lausanne (EPFL), Lausanne, Switzerland}
\affiliation[10]{Instituto de F\'isica Corpuscular, Centro Mixto Universidad de Valencia (CSIC), Valencia, Spain} 
\affiliation[11]{Instituto Galego de Física de Altas Enerxías (IGFAE), Universidade de Santiago de Compostela, Santiago de Compostela, Spain}
\affiliation[12]{Istituto Nazionale di Fisica Nucleare (INFN) Sezione di Genova, Genova, Italy}
\affiliation[13]{Istituto Nazionale di Fisica Nucleare (INFN), Sezione di Padova, Padova, Italy}
\affiliation[14]{Istituto Nazionale di Fisica Nucleare (INFN), Sezione di Pisa, Pisa, Italy}
\affiliation[15]{Istituto Nazionale di Fisica Nucleare (INFN), Sezione di Roma, Rome, Italy}
\affiliation[16]{LPNHE, Sorbonne Université, Paris Diderot Sorbonne Paris Cité, CNRS/IN2P3, Paris, France}
\affiliation[17]{Massachusetts Institute of Technology, Cambridge, MA, United States}
\affiliation[18]{Nikhef National Institute for Subatomic Physics, Amsterdam, Netherlands}
\affiliation[19]{Princeton University, Princeton, NJ, United States}
\affiliation[20]{Rice University, Houston, TX, United States}
\affiliation[21]{Rutgers University, New Brunswick, NJ, United States}
\affiliation[22]{School of Physics and Astronomy, University of Manchester, Manchester, United Kingdom}
\affiliation[23]{SUPA - School of Physics and Astronomy, University of Edinburgh, Edinburgh, United Kingdom}
\affiliation[24]{Tata Institute of Fundamental Research, Mumbai, India}
\affiliation[25]{Texas A\&M University, College Station, TX, United States}
\affiliation[26]{Universidade Federal do Rio de Janeiro (UFRJ), Rio de Janeiro, Brazil}
\affiliation[27]{Universit\`a degli Studi di Milano, Milan, Italy}
\affiliation[28]{Universit\`a di Pisa, Scuola Normale Superiore, and INFN Sezione di Pisa, Pisa, Italy}
\affiliation[29]{Universiteit Maastricht, Maastricht, Netherlands, associated to Nikhef National Institute for Subatomic Physics, Amsterdam, Netherlands}
\affiliation[30]{University of California, Riverside, CA, United States}
\affiliation[31]{University of California, Santa Barbara, CA, United States}
\affiliation[32]{University of Chinese Academy of Sciences, Beijing, China}
\affiliation[33]{University of Cincinnati, Cincinnati, OH, United States}
\affiliation[34]{University of Florida, Gainesville, FL, United States}
\affiliation[35]{University of Tennessee, Knoxville, TN, United States}
\affiliation[36]{University of Toronto, Department of Physics, Toronto, Ontario, Canada M5S 1A7}
\affiliation[37]{Van Swinderen Institute, University of Groningen, Groningen, Netherlands}
\affiliation[38]{Vinča Nuclear Research Institute, Vinča, Serbia}
\affiliation[39]{Vrije Universiteit Brussel, Brussels, Belgium}

\author[]{\\ \phantom{a} \\
The authors would like to thank everyone that worked on the ATLAS and CMS trigger system and software, as well as the LHCb Real Time Analysis (RTA) team for their useful feedback towards improving the contents and quality of this document. We also thank the LHCb computing, DPA, and simulation teams for their support and for maintaining the software on which our review of LHCb opportunities depend on.}

\author[]{\\ }

\abstract{
Long-lived particles (LLPs) are highly motivated signals of physics Beyond the Standard Model (BSM) with great discovery potential and unique experimental challenges. The LLP search programme made great advances during Run 2 of the Large Hadron Collider (LHC), but many important regions of signal space remain unexplored. Dedicated triggers are crucial to improve the potential of LLP searches, and their development and expansion is necessary for the full exploitation of the new data. The public discussion of triggers has therefore been a relevant theme in the recent LLP literature, in the meetings of the LLP@LHC Community workshop and in the respective experiments. This paper documents the ideas collected during talks and discussions at these Workshops, benefiting as well from the ideas under development by the trigger community within the experimental collaborations. We summarise the theoretical motivations of various LLP scenarios leading to highly elusive signals, reviewing concrete ideas for triggers that could greatly extend the reach of the LHC experiments. We thus expect this document to encourage further thinking for both the phenomenological and experimental communities, as a stepping stone to further develop the LLP@LHC physics programme. 
}

\begin{document}
\begin{flushright}
\hfill CERN-LPCC-2021-01
\end{flushright}
\maketitle

\newpage 
%%%%%%%%%%%%%%%%%

\section{Introduction and current gaps in coverage}
\label{s.introduction}
\emph{Section Editors: David Curtin, Nishita Desai, Jos\'e Zurita}
\vspace*{0.3cm}

In the last several years, long-lived particles (LLP)~\cite{Alimena:2019zri} have become a major focus in the search for new physics at the Large Hadron Collider (LHC), with the ATLAS~\cite{Aad:2008zzm}, CMS~\cite{CMS:2008xjf}, and LHCb~\cite{Alves:2008zz} experiments. LLP signals are generic in Beyond the Standard Model (BSM) scenarios, and specifically motivated in many theories that address fundamental puzzles like the hierarchy problem, baryogenesis, dark matter and the origin of neutrino masses~\cite{Curtin:2018mvb}. 
Run 2 of the LHC also brought great improvements in triggering on LLPs. To fully exploit the possibilities that the increased luminosity of Run 3 provides, triggering on LLPs should be preserved and expanded as new detector and data acquisition features become available\footnote{Technically unrelated to the trigger, but practically almost as impactful on LLP searches, are the data cleaning and reduction approaches undertaken by the experiments. Analysis datasets without out-of-time hits and displaced tracks can make an LLP search difficult or altogether impractical. As such, it would be highly beneficial to keep such considerations in mind as data management practices are revised or designed.}. In this document we summarise currently known gaps in experimental coverage with a special emphasis on gaps that arise due to trigger limitations, and review the ideas for possible new triggers at the three LHC experiments that may bridge these gaps. 

\subsection{Models with low \HT or displaced ``jets''}
\begin{itemize}
    \item {\bf Hidden Valley or dark-showers models} predict a secluded, i.e.\ ``hidden'' extended gauge sector with one or more mediators between the Standard Model (SM) and the new sector~\cite{Strassler:2006im}. The hidden sector typically consists of fermions which confine to form hidden mesons and baryons, which then can oscillate into SM mesons and decay visibly. However, the hidden hadrons could also be glueballs without any light hidden matter~\cite{Juknevich:2009ji}, which arises, for example, in theories of Neutral Naturalness~\cite{Chacko:2005pe, Craig:2015pha, Curtin:2015fna}.
    Production of dark fermions also results in radiation similar to parton showers in SM QCD~\cite{Carloni:2010tw}. Thus the final signature is of the form of emerging jets~\cite{Schwaller:2015gea} or Soft Unclustered Energy Patterns (SUEPs)~\cite{Knapen:2016hky}. 
    These models can easily evade \HT or jet-based triggers, especially if the dark shower is produced in exotic Higgs decays, which is the case in many motivated scenarios~\cite{Curtin:2013fra}, or if a large fraction of the energy does not oscillate back into the SM inside the detector.
    Moreover, the possibly unusual pattern of energy deposits (e.g., no tracks, unusual ratio of energy deposit in the electromagnetic calorimeter (ECAL) and hadronic calorimeter (HCAL), etc.) means they might not even be identified as jets.

    \item {\bf Light LLPs decaying hadronically} present another simple example of low-\HT scenarios. In the context of \HT or jet triggers, ``light'' can mean anything below the Higgs mass. Some of the simplest and most well-motivated examples of these scenarios are LLPs produced in exotic Higgs decays or other processes at or below the weak scale, which often decay via the Higgs portal (see e.g.~\cite{Curtin:2013fra}) and include solutions to the hierarchy problem like Neutral Naturalness~\cite{Chacko:2005pe, Craig:2015pha, Curtin:2015fna}; light pseudo-scalars that are predicted by extended Higgs sectors~(see e.g.~\cite{annurev-nucl-032620-043846} for a review); or axion models~\cite{Bauer:2017ris}. In such cases, a model-dependent trigger based on associated production of the light LLP with SM particles would currently be used. However, considerable improvement can be made if the products of the LLP can be used for software triggers -- often referred to as high-level triggers (HLT) -- such as displaced jets or leptons. LHCb can be used to particular advantage in the search for these models due to its exceptional ability to reconstruct secondary vertices and reconstruct light hadrons.

    \item {\bf LLPs with compressed spectrum} are predicted often by models of dark matter~\cite{Mahbubani:2017gjh, Fukuda:2017jmk, Filimonova:2018qdc, Bharucha:2018pfu} where the compression is a result of a physics requirement like obtaining the right relic density. Similar to prompt searches for compressed spectra, these models naturally suffer from low-\pt objects in the event and therefore need associated production of some other particles to be detected, often pushing the production cross section to undetectable levels~\cite{Blekman:2020hwr}. 
\end{itemize}

Several displaced jet or lepton triggers that build on the demonstrated success of existing triggers in both ATLAS and CMS~(see e.g.~\cite{Sirunyan:2020cao, Khachatryan:2014mea, CMS:2016isf, CMS_displacedLeptons_Run2, Aaboud_2019_trimuon}) could be implemented in Run 3 to dramatically improve sensitivity to these well-motivated BSM scenarios. 
Triggers based on exploiting higher energy deposition in the HCAL compared to the lower detector subsystems -- sometimes referred to as CalRatio-type triggers, which have already been used by ATLAS~\cite{Aad:2013txa,Aaboud:2019opc,Aad:2021cqq} -- or those that look for an absence of tracker hits could also be helpful in identifying signatures of some of these models in ATLAS and CMS. 
Finally, LHCb can contribute to cover a complementary region of the parameter phase-space in terms of light LLPs, due to its highly granular reconstruction and trigger capabilities.

\subsection{Models with displaced leptons or photons}
Triggering on displaced leptons offers a clean and model-independent probe of new LLPs that decay into leptons. Considerable improvement has already been made in Run 2 to take into account final states with hard displaced leptons~\cite{Aad:2020bay, CMS:2016isf,CMS_displacedLeptons_Run2}. However, several gaps still remain which can be itemised as follows:
\begin{itemize}
    \item {\bf Leptons with high impact parameter} are produced when a long-lived particle has lifetimes of the order of a few decimeters. This is a natural consequence when the coupling of the LLP to the SM is feeble, which occurs, for example, in models with freeze-in dark matter~\cite{Belanger:2018sti}. Models where the initial LLP is charged could in some cases be probed in charged-track searches instead. However, even in that case, the ability to reconstruct leptons from decays will be crucial in understanding the underlying theory. Current CMS searches for displaced leptons allow for a maximum offline displacement of 10~cm~\cite{CMS:2016isf,CMS_displacedLeptons_Run2} whereas the ATLAS search~\cite{Aad:2020bay} has an offline limit of 30~cm. Both of these could be improved by using information in the muon system. 

    \item {\bf Soft displaced leptons} provide another challenge to current lepton triggers which require at least $\pt \gsim 40$~\gev~\cite{Blekman:2020hwr}. Dark matter models with co-scattering~\cite{Filimonova:2018qdc, Bharucha:2018pfu} or models of compressed supersymmetry both often predict a peak in \pt-distribution at around 20~\gev or smaller. Furthermore, minimal assumptions (i.e.\ without extraneous assumptions on other parameters or new particles) predict production cross sections so low that a hard ISR mono-jet trigger (which currently is at $\sim 450$~\gev for both ATLAS and CMS) reduces the identification possibility to negligible levels.

    \item {\bf Leptons from dark photons or dark showers} will often show up as displaced vertices with only a few tracks~\cite{Buschmann:2015awa, Barello:2016zlb}, a signature sometimes referred to as ``lepton jets''. Current requirements on displaced vertices with a large number of tracks (e.g: $N_{track} \gtrsim 5$ ~\cite{Aaboud:2017iio, Sirunyan:2018pwn,CMS:2021tkn}) and a large invariant mass ($m_{DV} > 10$ GeV~\cite{Aaboud:2017iio}) of the vertex means that these events will be missed. Exclusive decays into two leptons are currently probed only when the \pt of the leptons is very high~\cite{Aad:2019tcc}. This can be particularly well remedied at LHCb which offers better vertexing and can be sensitive to much smaller lepton momenta. For example, the LHCb VELO used during Run 1 and Run 2 has a lifetime resolution of about 50~fs, an impact parameter resolution of 13--20 microns, and a secondary vertex precision of 0.01--0.05~mm in the xy plane~\cite{Borsato:2021aum}, while a “soft” track with \pt of about 1~GeV would have an impact parameter resolution of about 100 microns in CMS~\cite{CMS:2014pgm}. Furthermore, LHCb is able to reconstruct charged stable particles with $\pt > 80$ MeV~\cite{Borsato:2021aum}, as compared to 1 GeV for CMS~\cite{CMS:2014pgm}. 
    Moreover, since such dark shower models often predict multiple vertices (albeit of low mass), it may be possible to improve coverage by requiring multiple vertices with a small number of tracks in the same event, although this is a challenging signature with large backgrounds from pileup, for example.

    \item {\bf Models with displaced taus}~\cite{Desai:2014uha, Evans:2016zau} are currently nearly entirely unprobed except where they decay into (high-\pt) displaced electrons or muons~\cite{Aad:2020bay}. Since tagging tau decays is already complicated, it is not surprising that the tagging efficiency worsens when the tracks from the decay do not point to the primary vertex. 

    \item {\bf Displaced photons} are predicted by, e.g., radiative decay of the lightest neutralino to the gravitino in certain SUSY models like gauge mediation; see Ref.~\cite{Giudice:1998bp} for a review. A pair of displaced photons that do not point to the primary vertex can result from the decay of a new scalar or pseudo-scalar that is long-lived~\cite{Bauer:2017ris}. Combining a level-1 (L1) trigger on a prompt object along with HLT on displaced photons can improve the sensitivity to these models.
\end{itemize}

\subsection{Miscellaneous gaps}

If order of 100~ps timing information in the calorimeters and muon systems could be available at trigger level in ATLAS or CMS, then this new information would allow for more precise triggering on the products of LLPs or the LLPs themselves. Some possible applications are as follows:

\begin{itemize}
  \item {\bf Slowly moving LLPs} result when they are either produced from a very compressed decay of a particle at rest or due to the high mass of the LLP itself. They can be challenging to detect as they arrive in the outer detectors much later ($>1$~ns) than SM particles produced in the collision and moving at the speed of light, for which the current algorithms are optimised.

  \item {\bf Fractionally charged particles} can be difficult to detect because they do not deposit enough energy in the tracker or calorimeters to pass current trigger thresholds at tens of GeV. These thresholds may be reduced in the muon system using extra timing information.
\end{itemize}
Finally, we point out that using timing information will also allow reduction of thresholds on existing HLT cases, benefiting the low-\HT and low-\pt scenarios described above.

In the rest of this paper, we review experimental ideas of how to improve triggering on LLP decays at Run 3 of the LHC, and highlight the complementarity of approaches in ATLAS, CMS, and LHCb. The material for this document was to a large extent collected during several dedicated discussions in the framework of the LLP Community~(see e.g.~\cite{triggerLLPApr17, triggerLLPOct17, triggerLLPNov19, triggerLLPMay20, triggerLLPNov20}) and of the LHC LLP WG activities. Section~\ref{s.executiveSummary} gives a bird's eye overview and summary tables of possible new triggers that would mitigate the gaps outlined above. New opportunities for ATLAS and CMS are discussed in Section~\ref{s.ATLASCMS}, owing to the broad similarities of the two large detectors, though they are also discussed individually where appropriate.
LHCb differs substantially in its design and capabilities from the two larger detectors and offers several unique opportunities, which is why we discuss it separately in Section~\ref{s.LHCb}.

\vfill

\section{Executive summary of possible trigger improvements}
\label{s.executiveSummary}
\emph{Section Editors: David Curtin, Yuri Gershtein}
\vspace*{0.3cm}

%%%%%%%%%%%%%%%%%

%
Run 3 of the LHC promises to offer multiple opportunities for the discovery of new physics via increased luminosity and prominent changes---both hardware and software---to ATLAS, CMS, and LHCb.

Among exciting new possible developments for ATLAS and CMS are improvements to the hardware (L1) triggers, like increased segmentation and timing capabilities of calorimeter triggers and muon triggers without beam-spot constraints~\cite{CMS_Phase1_HCal,ATLAS_NewSmallWheel,ATLAS_Phase1_LArCal}. Progress in software and computing, as well as advances in machine learning (ML), can allow for better event reconstruction and more sophisticated analysis in the HLT. 

Changes specific to LHCb will also be significant, as there are plans to remove its hardware trigger altogether, allowing it be more sensitive than ATLAS and CMS for some signatures~\cite{Borsato:2021aum} despite lower luminosity and detector coverage. 
There are many untapped and new opportunities for LLP triggers and searches at LHCb, as the understanding of the BSM discovery potential of this detector has expanded considerably in recent years. Furthermore, the Run-3 upgrades remove the need for the hardware-based L0 trigger layer, allowing for full real-time event reconstruction and even more opportunities. 
Reflecting this capability, discussion of new triggers for LHCb is organised by the physical LLP decay mode into jets, exclusive hadron states, different leptons or photons. 
We emphasise the great complementarity between LHCb and ATLAS/CMS. LHCb could perform better for large regions of parameter space for BSM LLP signals at sub-100~\gev mass scales, since for such signals its enhanced triggering and tracking capability can more than make up for the reduced luminosity and geometric acceptance of its dataset. For larger mass scales, we expect better performance from ATLAS and CMS.

Since this document is intended to serve as a review of ongoing triggering techniques targeting LLP-associated final states, we organise the discussion by the particular experimental subsystem or capability that is most instrumental in developing the various new triggers.
We explicitly focus on the review of highly motivated trigger opportunities that are under development for the upcoming Run 3, since their development would be greatly beneficial for searches for unconventional signatures from new physics, which are often not reachable through standard triggering techniques.
Detector upgrades that are planned for the HL-LHC, including fundamentally novel tracking and timing capabilities at trigger level for ATLAS~\cite{CERN-LHCC-2017-020,Valente:2692161} and CMS~\cite{CMSCollaboration:2015zni,Tomei:2020wft}, and improved on-line reconstruction of low-momentum particles at LHCb~\cite{Aaij:2244311, Bediaga:1443882}, have the potential for even greater-reaching qualitative and quantitative impact on LLP searches.
%

%ATLAS and CMS
 \newcommand{\widthone}{2cm}
    \newcommand{\widthtwo}{4cm}
    \newcommand{\widththree}{4cm}
    \newcommand{\widthfour}{5cm}
    \newcommand{\widthfive}{1.4cm}

%ATLAS
\newcommand{\widthtwoA}{3.4cm}
\newcommand{\widthfourA}{4.7cm}
\begin{table}[hp]
    \centering
\hspace*{-1.5cm}
\begin{tabular}{|m{\widthtwoA}:m{\widthfive}|m{\widthtwoA}:m{\widthfive}|m{\widthfourA}|m{\widthfive}|}

   \hline
    \textbf{L1} & \textbf{Present in Run 2} & \textbf{HLT} & \textbf{Present in Run 2} & \textbf{Physics motivation example} & \textbf{Section} \\
   \hline \hline 

   Jet or MET & Yes & * Number of tracker hits ``below'' jet & No & Hadronically decaying LLPs with low-HT where displaced track reconstruction is particularly difficult  & \ref{s.displacedjetsatlas}
    \\ \hline 

   * HCAL timing & No & Various & Yes & Slow LLPs (heavy or produced near threshold) & \multirow{2}{\widthfive}{\ref{s.newL1usingcalorimeters}}
   \\
   \cline{1-5} 
   * HCAL timing + CalRatio type & No & Various & Yes &  LLPs decaying in calorimeter &
   \\ \hline

   1$j$, 3$j$, \HT, 2$\tau$ & Yes & * Calo timing (+ tracking?) + dramatic reduction of HLT thresholds & No &  Various LLP scenarios   &   \multirow{2}{\widthfive}{\ref{s.calotiming}}
    \\
    \cline{1-5}
    Photon & Yes & Displaced $\gamma$ + * timing & No & GMSB & 
   \\ \hline

    \vspace{2mm}Single muon\vspace{3mm} & Yes & \multirow{3}{\widthtwoA}{*Displaced track(s) in inner detector (*add calo timing for electrons?)} & \multirow{3}{\widthfive}{No} & 
   \multirow{2}{\widthfourA}{Soft displaced leptons; GMSB staus, freeze-in DM, LLPs from Higgs boson decays}  &
  \multirow{3}{\widthfive}{\ref{s.singledisplacedleptons}}
   \\
   \cline{1-2}
    \vspace{2mm}Single electron\vspace{3mm} & Yes  & & & &  \\
   \cline{1-2} \cline{5-5}
     Di- (or tri-) muon & Yes & & & Soft displaced multi-lepton, e.g.\ dark photons, dark shower & 
   \\ \hline

   Muon system & Yes & * Muon system timing & No & Fractionally charged particles & \ref{s.fcp}
   \\ \hline
    * Displaced muons & No & Muon system and inner tracker & Yes & Displaced muons with impact parameter $>$ 10s of cm
   & \ref{s.muonatlas}, \ref{s.muoncms}
    \\ \hline

   \end{tabular}
    \caption{
    Summary of ideas for new Run-3 triggers for ATLAS.
    We assume that Run-2 triggers~\cite{Aad:2021cqq,ATLAS:2016wtr} will be retained or improved for Run 3.
    The new component of each trigger is marked with a star *. Question marks indicate possibilities that need further investigation. Please refer to text for further details.    
    }
    \label{tab:atlassummary}
\end{table}

%CMS

\newcommand{\widthtwoB}{3.4cm}
\newcommand{\widthfourB}{4.7cm}
\begin{table}[hp]
    \centering
    \vspace*{-0.5cm}
\hspace*{-1.5cm}
\begin{tabular}{|m{\widthtwoA}:m{1.4cm}|m{\widthtwoA}:m{1.4cm}|m{\widthfourA}|m{1.1cm}|}

   \hline
    \textbf{L1} & \textbf{Present in Run 2} & \textbf{HLT} & \textbf{Present in Run 2} & \textbf{Physics motivation example} & \textbf{Section} \\
   \hline \hline 
   
   di-tau &  Yes & 
   \multirow{2}{\widthtwoA} {* Displaced jet with low \HT, working off L1 seed} & No
  & LLPs in low-\HT events, e.g.\ from exotic Higgs decay 
  & \multirow{3}{\widthfive}{\ref{s.displacedjetscms}}
   \\ 
   \cline{1-2} \cline{5-5}
    other $\gamma$/$\ell$/MET with L1 threshold $<$ HLT threshold & Yes & & & LLPs in low-\HT events, also emerging jets, dark showers, disappearing tracks, etc. & 
   \\ 
   \cline{1-5}
    hard jet/lepton/photon & Yes & 
   * Displaced jet \emph{opposite} to hard object & No 
   & displaced jet + X searches, e.g.\ low mass LLP or dark shower recoiling against SM objects& 
   \\ \hline

   Jet or MET & Yes & * Number of tracker hits ``below'' jet & No & Hadronically decaying LLPs with low-HT where displaced track reconstruction is particularly difficult  & \ref{s.displacedjetsatlas}
    \\ \hline 

   * HCAL timing & No & Various & Yes & Slow LLPs (heavy or produced near threshold) & \multirow{2}{\widthfive}{\ref{s.newL1usingcalorimeters}}
   \\
   \cline{1-5} 
   * CalRatio type (analogous to existing ATLAS capability) & No & Various & Yes &  LLPs decaying in calorimeter &
   \\ \hline

   1$j$, 3$j$, \HT, 2$\tau$ & Yes & * Calo timing (+ tracking?) + dramatic reduction of HLT thresholds & No & Various LLP scenarios & \multirow{2}{\widthfive}{\ref{s.calotiming}}
    \\
    \cline{1-5}
    Photon & Yes & Displaced $\gamma$ + * timing & No & GMSB & 
   \\ \hline

    \vspace{2mm}Single muon\vspace{3mm} & Yes & \multirow{3}{\widthtwoA}{*Displaced track(s) in inner detector (*add calo timing for electrons?)} & \multirow{3}{\widthfive}{No} & 
   \multirow{2}{\widthfourA}{Soft displaced leptons; GMSB staus, freeze-in DM, LLPs from Higgs boson decays} &
  \multirow{3}{\widthfive}{\ref{s.singledisplacedleptons}}
   \\
   \cline{1-2}
    \vspace{2mm}Single electron\vspace{3mm} & Yes & & & & \\
   \cline{1-2} \cline{5-5}
     Di- (or tri-) muon & Yes & & & Soft displaced multi-lepton, e.g.\ dark photons, dark shower & 
   \\ \hline

   Muon system & Yes & * Muon system timing & No & Fractionally charged particles & \ref{s.fcp}
   \\ \hline
    * Displaced muons & No & Muon system and inner tracker & Yes & Displaced muons with impact parameter $>$ 10s of cm
   & \ref{s.muonatlas}, \ref{s.muoncms}
    \\ \hline
   
 * Hadronic LLP decay in muon system (analogous to existing ATLAS capability) & No & * Using information from all detector layers near the muon ROI & No & Hadronic LLPs decaying in muon system & \ref{s.muoncms}
    \\ \hline
 
   \end{tabular}
    \caption{
    Summary of ideas for new Run-3 triggers for CMS.
    We assume that Run-2 triggers~\cite{CMS:2020cmk,Khachatryan:2016bia,Sirunyan:2021zrd} will be retained or improved for Run 3.
    The new component of each trigger is marked with a star *. Question marks indicate possibilities that need further investigation. Please refer to text for further details.    
    }
    \label{tab:cmssummary}
\end{table}

  \renewcommand{\widthone}{1.5cm}
    \renewcommand{\widthtwo}{6cm}
    \renewcommand{\widththree}{7cm}
    \renewcommand{\widthfour}{1.2cm}
\begin{table}[hp]
    \centering
\hspace*{-1.5cm}
    \begin{tabular}{|m{\widthtwo}|m{\widththree}|m{\widthfour}|}

   \hline
    \textbf{New LLP Trigger} & \textbf{Physics motivation example} & \textbf{Section} \\

   \hline \hline

     Displaced fat jet at HLT2 & 
Hadronically decaying LLPs $<$ 15~\gev
    & \multirow{2}{\widthfour}{\ref{s.lhcbdisplacedjets}}
    \\
    \cline{1-2}
     Displaced vertices at HLT2 with high track/vertex multiplicity & General LLPs that decay hadronically 
    & 
    \\ \cline{1-3}

New topological LLP trigger based on existing exclusive trigger for $B$-decays. 
 & 
Hadronically decaying LLPs with masses and lifetime below that of $B$-meson
    & \multirow{2}{\widthfour}{\ref{s.displacedhadronslhcb}}
    \\
    \cline{1-2}
     Triggering on exclusive LLP $\to$ $K^+K^-, \pi^+ \pi^-$, etc. decays at HLT2 & Hadronically decaying LLPs with masses $\sim \mathcal{O}(\gev)$
    & 
    \\ \cline{1-3}

Displaced di-tau at HLT2
 & 
LLPs that decay through Higgs portal, in particular for $\lesssim$ 50~\gev masses that are most challenging for main detectors
    & \multirow{2}{\widthfour}{\ref{s.displacedditaulhcb}}
    \\
    \cline{1-2}
     
Prompt tau + displaced tau at HLT2
 & 
Heavy neutral leptons
    & 
    \\ \cline{1-3}

Soft di-electron without L0 limitation
 & 
Dark photon, dark showers
    & \multirow{2}{\widthfour}{\ref{s.displaceddiphotonselectronslhcb}}
    \\
    \cline{1-2}
     
Soft di-photon without L0 limitation
 & 
Axion-like particles, dark showers
    & 
    \\ \cline{1-3}

 Downstream tracks (hits in every tracker except VELO)
&
LLPs with decay length $\gtrsim$ 10s of cm
&
\vspace*{2mm}

 \ref{LHCb.downstream}

    \\ \cline{1-3}
      
Utilising RICH sub-detectors to trigger on production of anomalous Cherenkov photons
 & 
Fractionally charged and massive stable charged particles
    & \multirow{2}{\widthfour}{\ref{s.lhcbchallengingideas}}
    \\
    \cline{1-2}
     
Cluster occupancies of ECAL and HCAL, track event isotropy
 & 
SUEPs
    & 
\\ \hline
   
  \end{tabular}
    \caption{Summary of new LHCb trigger ideas for Run 3. Please refer to the text for further details. 
    }
    \label{tab:lhcbsummary}
\end{table}

\clearpage

Although possible signatures with LLPs are very diverse, it is of utmost importance to improve triggering on low energy sum (\HT) events with soft displaced jets or leptons, which have low acceptance in existing searches, such as Ref.~\cite{Sirunyan:2020cao}. For example, for $p p \to a a$, with the mass of $a$ equal to 20 GeV, lowering the \HT cut from 500 GeV to 150 GeV will result in a a factor of 10 increase in signal. As discussed in the previous section, events like these could arise in a plethora of theoretical scenarios. Throughout this review of new trigger developments, it is helpful to consider some specific benchmark models that can be used to demonstrate the physics case for the suggested capability. 
One possibility is the decay of the Higgs boson, which could be the only portal to the dark sector accessible at the LHC. Another is the electroweak production of supersymmetric partners in case of $R$-parity violation, low-scale gauge-mediated supersymmetry (SUSY) breaking, or if the lightest SUSY particle is in the dark sector.
Dark photons, dark showers, axions and heavy neutral leptons are also highly motivated, either in their own right or as part of more complete models, and frequently lead to such LLP signatures.

In Tables~\ref{tab:atlassummary} and \ref{tab:cmssummary}, we present a review of the new ATLAS and CMS trigger ideas for Run 3 of the LHC, many of which were developed by the ATLAS and CMS Collaborations.
The ideas reported here are explored at both main detectors, though some are tailored to specific capabilities or gaps in coverage of one of the experiments.
%nt
Table~\ref{tab:lhcbsummary} shows the new triggers ideas for LHCb. The versatility of the upgraded trigger system means that there are many more advanced trigger possibilities utilising the GPU at HLT1 that have to be investigated, and we refer the reader to the main text for detail.

Implementation of these triggers would greatly enhance the discovery potential of the LHC, taking full advantage of the complementary capabilities of ATLAS, CMS and LHCb in searching for BSM LLP signals.

\section{ATLAS and CMS}
\label{s.ATLASCMS}

There are several generic strategies to improve trigger efficiency for long-lived signals at ATLAS and CMS which inform our discussion. 
One of the simplest and most important approaches is to require a long-lived component at the HLT in combination with almost any L1 seed.
This strategy is most useful when there is a large gap between HLT and L1 thresholds. The long-lived component can be used to reduce background rates, and allow HLT thresholds to be closer to L1 levels. Most often, this long-lived requirement makes use of tracking information which was not available at L1. For example, the CMS Run-2 displaced-jet trigger reduces the \HT threshold from 1~\tev to 430~\gev by requiring jets have at least one slightly displaced track, and at most two prompt tracks~\cite{Sirunyan:2020cao}.

Cross triggers represent another strategy which can help reduce \pt thresholds~\cite{CMS:2020cmk}. Cross triggers require multiple objects in order to improve background rejection, at the cost of creating a more model-dependent trigger selection. This strategy is most useful for models which predict multiple soft objects. In Run 2, cross triggers have been used to target compressed scenarios in prompt SUSY models, which result in soft leptons and relatively low transverse momentum imbalance (\met)~\cite{CMS:2018kag,CMS-DP-2020-004}.

Trigger-level analysis~\cite{ATLAS:2018qto} / data scouting~\cite{CMS:2016ltu} as well as delayed streams~\cite{ATLAS:2014ktg} / data parking~\cite{cms_parking} are also potentially interesting strategies to improve trigger efficiency for long-lived particles at ATLAS and CMS, which might help implement any of the ideas discussed in this work.
Data scouting allows for substantially reduced \pt thresholds, and correspondingly larger event rates, by only saving partial results of the online event reconstruction. The full event information for scouting data is not available, so it is not possible to take advantage of the improvements offline calibration and reconstruction provide.
However, since any partial event information that is available to the HLT can be saved, this can include tracker information that may be particularly useful for displaced decays.
It would therefore be interesting to consider what kind of data scouting techniques might increase reach for LLP scenarios. 
Data parking represents a complementary strategy by using triggers with lower thresholds to save the full event content, yet reconstructing physics objects only during shutdowns when extra computing resources become available. 

%%%%%%%%%%%%%%%%%
\subsection{Using tracker information at the HLT for displaced jets}
\label{s.displacedjets}
%%%%%%%%%%%%%%%%%
\emph{Section Editors: 
Matthew Citron, David Curtin, Emma Torro}
\vspace*{0.3cm}

In this section we explore using tracker information at the HLT, seeded by a variety of L1 triggers, to significantly increase acceptance for LLP signals decaying in the tracker or calorimeter.
Using tracks or tracker hits in LLP triggers has the potential to significantly improve sensitivity to hadronically decaying LLPs, which suffer from high trigger thresholds and quality requirements in the absence of leptons or copious \met, as well as other LLP scenarios that have low efficiencies for existing triggers.

\subsubsection{CMS}
\label{s.displacedjetscms}

In Run 2, a dedicated displaced-jet trigger allowed the threshold on the total hadronic energy in the event to be reduced from over 1~\tev to the L1 plateau of $\sim 450$~\gev~\cite{Sirunyan:2020cao}. 
This allowed hadronically decaying $R$-parity violating (RPV) SUSY superpartners or LLPs arising from scalar decay to be constrained with masses as low as $\mathcal{O}(100~\gev)$.
Importantly, the resulting sensitivity to LLPs from exotic decays of the 125~\gev Higgs boson is world-leading for this decay mode and lifetimes below a meter, even though the L1 \HT trigger threshold of 360 GeV~\cite{CMS:2020cmk} severely reduces sensitivity for signal mass scales $\lesssim 100~\gev$.

In Run 3, the displaced jet HLT would be greatly improved by being seeded with a wider variety of general-purpose L1 triggers. 
The most obvious candidate for another L1 trigger to seed displaced-jet reconstruction at the HLT are di-taus, which have a much lower threshold than the multi-jet triggers. The narrower cone used in the tau reconstruction would fit the signal topology of an LLP undergoing a displaced decay, not only to hadronic $\tau^+ \tau^-$ but also $b \bar b$ and light-flavour di-jets. 

LLPs interacting with the SM via the Higgs portal, such as glueballs in Neutral Naturalness~\cite{Chacko:2005pe, Craig:2015pha, Curtin:2015fna}, often decay with Yukawa-ordered branching ratios. 
This means that decays to $\tau \tau$ and $\bar b b$ are dominant. Given the low efficiency for LLPs from Higgs decays to pass the L1 \HT threshold (using Run-2 assumptions), running the displaced-jet high-level trigger off tau L1 seeds has the potential to be the best strategy for LLPs produced in exotic Higgs-boson decays in significant regions of parameter space. 

The hadronic nature of the shower or the possible non-projectiveness might interfere with isolation requirements placed on the low-\pt di-tau triggers. Experiments (both ATLAS and CMS) are therefore studying the LLP signal inefficiencies due to isolation, and might also need to consider non-isolated tau triggers with higher \pt thresholds. 

Single-tau and single-jet triggers have similar thresholds at L1. A tau-specific triggering strategy for displaced jets is therefore unlikely to provide significant improvement in the case of single displaced taus.

The utility of displaced-jet high-level triggers to reduce thresholds to the L1 plateau is completely general as long as there is a significant gap between the non-displaced L1 and HLT thresholds. This has the potential to open up new sensitivity for moderate- or low-mass LLPs below several hundred \gev for a variety of models. 
Triggers that match the displaced jet at HLT to the L1 photon, muon or \met are therefore highly motivated for exploration in Run 3. For example, tracking to reconstruct emerging jet or dark shower signatures~\cite{Schwaller:2015gea, Alimena:2019zri}, which can involve a significant number of displaced decays to multiple objects within a jet, could be seeded by such triggers. In addition, a disappearing track signature (recoiling against ISR) could be reconstructed in the vicinity of the \met. 

Another option is to adapt the ``mono-X + MET'' search strategy for dark matter simplified models to look for LLPs, but requiring displaced jets instead of \met.
Specifically, the presence of a jet, lepton, or photon of sufficient momentum at L1 could potentially be used to seed a more displaced-tracking iteration at the HLT in a small region \emph{opposite} to the high-momentum object.
This could then, in principle, be used to enhance efficiency at HLT for relatively light LLPs that can recoil an associated prompt SM object. 

All of the above techniques could incorporate machine-learning methods at HLT, similar to those used offline in Run 2~\cite{Sirunyan:2020cao}, to improve signal efficiency.

\subsubsection{ATLAS}
\label{s.displacedjetsatlas}
Neutral LLPs decaying in the calorimeter give a very distinctive signature. The ATLAS detector in Run 2 featured one HLT selection path dedicated to triggering on displaced jets appearing in the calorimeter~\cite{Aaboud:2019opc}. It was based on two main characteristics: the lack of tracks associated to the jet and a small fraction of the energy deposited in the electromagnetic calorimeter.
Two L1 triggers were used as seeds to this selection. The lowest unprescaled L1 tau seed had very good efficiency for displaced jets with $\pt > 100$~\gev. For lower \pt, a dedicated L1 topological selection~\cite{Aad:2021cqq} was designed, selecting L1 jets with all their energy deposited in the HCAL. This tight selection on the electromagnetic fraction allowed for a lower \pt threshold, with a plateau at 60~\gev.

Retaining and improving these triggers in Run 3 will be essential, and improving them would be highly beneficial. For example, the fraction of energy deposited in every calorimeter layer could potentially be used for a better signal--background separation. Additionally, ML techniques have been used offline~\cite{Aaboud:2019opc} with positive results and the same techniques could be exported to the HLT. Improvements in pile-up rejection and in-track isolation are also highly significant for Run 3.

Additional triggers that target hadronic decays of LLPs in the inner detectors (IDs) could be very beneficial. For example, while ATLAS has triggers that target decays in the calorimeters or in the muon system, there is currently no dedicated trigger for decays in the ID. Such a trigger could potentially greatly increase sensitivity to short-lived LLPs, e.g., those that decay somewhere in the middle of the ID~\cite{Triggering_EmergingJets}. Such triggers have not been developed in the past due to the difficulty of performing high-$d_0$ tracking at the HLT.
Developing such a capability at the HLT would allow ATLAS to significantly expand their capabilities searching for LLP decays in the tracker, so new ideas and approaches here would be very welcome.

One option could be to use the existing tracking capabilities within the HLT system to simply count ID hits not associated to standard tracks, as the expected signature of a hadronic decay of an LLP in the ID would be a high multiplicity of hits appearing a certain three-dimensional distance from the interaction point (IP), largely grouped together in $\Delta R$, with few to no ID hits between the IP and the beginning of the hit cluster~\cite{Triggering_EmergingJets}.
Such a trigger could potentially be seeded by standard L1 jet or \met\ seeds, and then hits could be counted within a cone around a region of interest (RoI) defined by a calorimeter jet or another RoI. Moreover, dedicated seeds could be developed if existing seeds are inefficient. One can also imagine a regional tracking iteration geared towards reconstruction of tracks originating in the small volume at a radius just before the tracking layer with large hit multiplicity.
%

%%%%%%%%%%%%%%%%%
\subsection{Using calorimeter-based LLP triggers at L1 and HLT}
\label{s.calo.timing}

\emph{Section Editors: 
Javier Montejo Berlingen, Matthew Citron}
\vspace*{0.3cm}

The offline time resolution of the ATLAS and CMS calorimeter cells is well below 1~ns, which makes it an excellent discriminant to identify out-of-time energy deposits from slow-moving particles. The ATLAS calorimeters have an offline time resolution of about 300~ps and 600~ps for energetic cells in the ECAL and HCAL respectively~\cite{Aad:2014gfa,Davidek_2017}. CMS shows a similar performance offline with 400~ps and 1~ns in the ECAL and HCAL respectively. 
A recent search by CMS for non-prompt jets~\cite{2019134876} defines the signal region with at least one jet with time $> 3$~ns. 
This search relies on \met triggers and places stringent offline requirements on it ($\met > 300$~\gev) in order to reduce the amount of background. 
This cut makes the search insensitive to signals with a long-lived particle with mass under $\sim 1$~\tev and for models such as gluino--bino coanihillation in which there is a small mass splitting~\cite{Nagata:2015hha,Nagata:2017gci}.

As a result, there is a strong need for triggers that select jets with delays of about 1~ns, which would allow dedicated searches in previously unexplored phase space, as well as triggers with more stringent timing selections but reduced energy thresholds to target models with compressed spectra.
This can proceed on two fronts: development of new L1 triggers using calorimeters, and use of calorimeter timing at HLT with a variety of L1 seeds. 
CMS is also exploring the depth-based L1 triggering capability deployed by ATLAS to search for LLPs decaying in the hadronic calorimeter~\cite{CMSHcalLLPs}.

\subsubsection{New L1 triggers using calorimeters}
\label{s.newL1usingcalorimeters}

Using calorimeter timing information at L1 has the potential to significantly reduce energy thresholds for delayed-jet signals if the information is available and precise enough~\cite{CMSHcalLLPs}. Such a strategy has already been proposed for Phase~2~\cite{CERN-LHCC-2020-004,Bhattacherjee:2020nno} and could also be explored for Run 3. 
For example, the CMS HCAL may be expected to provide a timing measurement on the order of a few ns at L1~\cite{CMS_LLP_triggers_for_Run3}. %~[some ref]. \DC{ref}
Even with a suboptimal time resolution, it would allow triggering on very slow moving particles. However, several of the subsystems might not have timing information available, not have enough bandwidth to send it for use outside the firmware, or have poor timing resolution. Dedicated studies to determine if such triggers are viable has been carried out for the CMS Phase-2 Upgrade L1 TDR, considering Higgs bosons decaying into pseudoscalars decaying into displaced jets. RPV SUSY is another relevant benchmark to consider, and dedicated L1 seeds could be implemented already at Run 3. Delays on the order of a few ns would bring new sensitivity~\cite{2019134876,Liu:2018wte}.

Triggers that rely on energy imbalance between HCAL and ECAL have been successfully deployed by ATLAS~\cite{Aad:2013txa,Aaboud:2019opc} in order to collect events containing decays that occur inside the HCAL. Such triggers could be developed for CMS as well. In Run 3, the CMS HCAL will be able to be read out in three segmentation depths for the HCAL barrel and seven depths for the HCAL end-cap, enabling the reconstruction of energy deposits in the transverse direction. The depth information can, in principle, be exploited to further reduce the thresholds for a decay within the calorimeter.
For both ATLAS and CMS, combining depth and timing information could further lower thresholds or loosen the quality and isolation criteria for timing-based triggers. 

\subsubsection{Exploiting calorimeter timing at HLT}
\label{s.calotiming}
All cell information is available at the HLT, and the jet time, defined as the median cell time, could be easily computed. The hadronic triggers at ATLAS and CMS in Run 2 were limited by the HLT output rate. As a result, requirements on the jet timing could in principle be used to reduce thresholds significantly~\cite{Liu:2018wte}, down to the L1 plateaus~\cite{CMS_LLP_triggers_for_Run3}. The improvements could be substantial for several triggers, including single-jet triggers (for example, from an offline HLT jet momentum threshold of $\pt > 450$~\gev down to the L1 level of $\pt > 200$~\gev for ATLAS)~\cite{ATLAS_jet_trigger}, multi-jet triggers, and \HT triggers (for example, from $\pt > 1$~\tev down to $\pt \sim 400$~\gev for CMS)~\cite{Sirunyan:2020cao,Sirunyan:2019xwh}. To allow such dramatic reductions in thresholds will likely require the use of tracking information in addition to timing (see Section~\ref{s.displacedjets}). 
In general, high-level software triggers would most likely greatly benefit from exploiting both timing- and depth-based seeds at L1, potentially utilizing the HCAL and ECAL energy ratios as well as highly displaced tracking iterations in the region near the jet.
Such algorithms could also be seeded by non-isolated hadronic tau triggers at L1 (see Section~\ref{s.displacedjetscms}). 

In addition to hadronic triggers, timing at HLT may also be used to trigger displaced photons, which can be produced in the decay of long-lived neutralinos in gauge-mediated supersymmetry breaking (GMSB). In Run 2, CMS used a trigger consisting of a non-pointing photon of $\pt > 60$~\gev in association with $\HT > 350$~\gev~\cite{Sirunyan_2019}. The use of timing for the displaced photon at HLT could be investigated in Run 3, as it could potentially allow the \HT threshold to be significantly reduced or removed, widening the acceptance of the search to GMSB scenarios and other models with lower hadronic activity~\cite{CMS_LLP_triggers_for_Run3}. 

Timing-based triggers will face non-standard backgrounds in addition to mismeasured QCD. Such backgrounds include satellite-bunch collisions (collisions of low-luminosity bunches at 2.5~ns spacing from the IP), energy deposits from beam-halo muons, and energy deposits from cosmic-ray muons~\cite{2019134876}. 
To properly estimate these contributions in an offline analysis it may be necessary to devise additional triggers to collect background-enriched 
samples. 
Satellite-bunch collisions may form a useful out-of-time sample to evaluate the reconstruction of delayed energy deposits. If complementary information from the tracker is used to identify displaced or trackless jets, an additional trigger without these requirements (using a higher energy threshold, tighter isolation or quality criteria, or a prescale) will most likely be needed to collect such a sample.

Finally, it should be noted that, currently, out-of-time deposits are penalised in their energy reconstruction in order to mitigate out-of-time pileup~\cite{Sirunyan_2020_multifit}. For significantly delayed deposits this can cause inefficiency in reconstruction. While this effect was reasonably small in Run 2, tighter cuts could severely limit future sensitivity and should therefore be avoided in Run 3. Alternatives that more precisely reject the narrow time windows (at 25~ns intervals) of out-of-time pileup could also be explored.

%%%%%%%%%%%%%%%%%
\subsection{Using lepton L1 seeds to enable tracking of displaced leptons at the HLT}
\label{s.displacedleptons}
%%%%%%%%%%%%%%%%%
\emph{Section Editors: 
Freya Blekman, Tova Holmes, Karri Di Petrillo}

 Low-\pt triggers for leptons with displacements on the order of a millimeter would be very beneficial for ATLAS and CMS. A potential strategy to accomplish this would involve using standard L1 lepton triggers to seed HLT Inner Detector (ID) tracking. By requiring leptons to have a well-matched, slightly displaced ID track, it may, in principle, be possible to keep the lepton \pt requirements closer to L1 thresholds. 

This strategy should already be feasible for both electrons and muons. In both cases, it would be beneficial to extend the efficiency of any regionally-seeded displaced tracking as far as possible in lepton impact parameter. A careful optimisation of the minimum impact parameter required in order to keep the lepton \pt threshold as low as possible would also likely be vitally useful.

\subsubsection{Single displaced leptons}
\label{s.singledisplacedleptons}
Single prompt-lepton triggers for both ATLAS and CMS are initiated by L1 seeds based on calorimeter or muon spectrometer (MS) information, which operate at a \pt threshold of about 25~\gev~\cite{ATL-DAQ-PUB-2018-002}. At the HLT in ATLAS, region of interest (RoI)-based ID tracking is incorporated. For electrons in CMS at the HLT, the tracking is an-ECAL driven or ``outside-in'' based tracking, where the algorithm searches for pixel seeds that are compatible with the supercluster trajectory and then forms a seed for the ``GSF Tracking'' from 2 or 3 pixel hits, depending on the number of layers crossed. Requiring leptons to have a well-matched, isolated, ID track keeps the threshold for prompt lepton triggers close to the L1 threshold. By comparison, HLT chains without ID track requirements are much more rate-constrained. On ATLAS, the single-muon trigger without an ID track constraint has a threshold at 80~\gev, while the electron equivalent (a single-photon trigger) has a threshold at 140~\gev.

ATLAS and CMS could both benefit from a new trigger designed for single displaced leptons which keeps the HLT threshold closer to the L1 threshold by additionally requiring a slightly displaced ID track. In the ATLAS HLT, tracks are reconstructed with nearly full efficiency up to impact parameters of $\sim 10$~mm~\cite{Grandi:2728111}. In the CMS HLT, the tracking efficiency is also nearly fully efficient at $\sim 10$~mm, and drops to about $\sim 50\%$ at impact parameters of 30~mm~\cite{Tosi:2016bat}. It might be possible to extend this reconstruction efficiency slightly with a regionally-seeded, displaced ID tracking step.

Signals with single soft leptons would greatly benefit from such a trigger. One example of such a signature would be long-lived particles which decay to a tau and an invisible particle. This signature is predicted in GMSB stau scenarios, lepton-flavoured dark matter and freeze-in dark–matter models~\cite{Evans_2016}.
With two taus in the final state, 45\% of events would contain one light lepton and one hadronically decaying tau. Light leptons from the displaced tau are typically soft (with transverse momentum roughly scaling as $\pt \sim \frac{1}{6}$ of the long-lived particle mass) and depending on the specifics of the phase space, often have very little trigger efficiency with the current trigger thresholds of 50--80~\gev. Decays of the SM-like Higgs boson to long-lived particles serve as another interesting motivation. Several models predict significant branching ratios to taus, resulting in a large fraction of events with at least one soft lepton.

\subsubsection{Multi-lepton triggers}

Multi-lepton triggers represent another avenue to reduce \pt thresholds with respect to single-lepton triggers. A wide variety of long-lived particle models result in two displaced leptons in the final state. These models include scenarios in which the leptons are produced from the same decay or from separate long-lived particle decays. 

During Run 2, CMS had several triggers designed to target events with two displaced leptons~\cite{Khachatryan:2014mea, CMS:2016isf,CMS_displacedLeptons_Run2}. These triggers selected $ee$, $e\mu$ and $\mu\mu$ events with a $\pt$ threshold on each lepton of 30--50~\gev, and no primary vertex requirement. A trigger requiring a photon and a displaced muon was used to select events with a displaced muon and displaced electron, or the equivalent with two displaced electrons (i.e., trigger on two photons). The thresholds for these triggers were substantially increased when the performance of the LHC increased during Run 2. To achieve low-\pt thresholds during Run 3 further work will be necessary. 

In Run 2, ATLAS did not have corresponding di-lepton triggers targeting displaced light leptons. As a result, for ATLAS in Run 3, a di-standalone-muon trigger and a trigger that includes a standalone muon and a photon represent potential avenues to trigger on displaced light leptons with reduced \pt thresholds.

Both ATLAS and CMS could potentially further reduce the \pt thresholds of these displaced di-lepton triggers by additionally requiring one or both leptons to have a well matched, slightly displaced ID track. Such a strategy could potentially reduce HLT \pt thresholds to as low as the L1 floor of $\sim$ 10~\gev and could potentially improve performance for both electrons and muons.

ATLAS and CMS have also previously used prompt tri-muon triggers with low \pt thresholds that do not have requirements on displacement. For ATLAS the prompt \pt thresholds varied from 4 to 6~\gev throughout Run 2, with HLT thresholds very close to L1, and there was no isolation requirement. ATLAS also had a tri-muon trigger using only standalone MS information, which can select displaced leptons with $\pt > 6$~\gev~\cite{Aaboud_2019_trimuon}. As with other displaced-muon triggers, this threshold could likely be reduced with the addition of displaced ID track requirements at HLT. Thresholds lower than 4~\gev would most probably require changes to the L1 seeds, and are unlikely to be possible.

Soft, displaced multi-muon triggers may also potentially provide significantly increase reach for, e.g.,\ low-mass dark photon models.
For dark shower models which create a large number of low-\pt muons, these small improvements in \pt threshold for multi-muon triggers could also have a very significant impact~\cite{Alimena:2019zri}.

\subsubsection{Other lepton-based trigger strategies}

Cross triggers represent an attractive strategy for models that predict single displaced leptons produced with additional SM particles or \met. Such scenarios are common in supersymmetric models. The prospect of requiring a slightly displaced, low-\pt lepton to ensure that \met or other thresholds are kept close to L1 levels and that signal efficiency is greatly improved, could be explored. Other strategies to reduce the trigger rate for di-lepton triggers, including machine learning techniques that rely on specific signal models, were recently proposed in Ref.~\cite{Blekman:2020hwr}.

Electron triggers traditionally are subject to more challenging backgrounds, and so often require tighter L1 and HLT \pt thresholds or tighter isolation or quality requirements. There is further potential for improvement for the reconstruction of displaced electrons if L1 photon objects associated with displaced tracks could be combined with the strategies used to select delayed calorimeter objects discussed in Section~\ref{s.calotiming}.

Data scouting and data parking represent another potential avenue to access low-\pt displaced leptons. In Run 2, CMS was able to use scouting to select events with two muons, with an invariant mass above 200~\mev and a muon momentum above 3~\gev, with and without a primary vertex constraint~\cite{Duarte:2018bsd,Mukherjee:2017wcl,Anderson:2016ron,CMS-DP-2018-055,CMS-DP-2021-005}. Such a strategy would be useful if an analysis could be performed with limited event content. CMS was also able to park nearly 10 billion b-hadron pairs by triggering on events with slightly displaced muons. This strategy could look for long-lived particles in the unbiased b-decay, or such a strategy could be adapted for other long-lived scenarios.

\vspace*{0.3cm}

%%%%%%%%%%%%%%%%%
\subsection{Using L1 muon seeds to enable use of muon-system timing information at the HLT}
\label{s.muontiming} 
%%%%%%%%%%%%%%%%%
\emph{Section Editors: 
Karri Di Petrillo}
\vspace*{0.3cm}

Both ATLAS and CMS have muon detectors with per-hit time-of-arrival resolution on the order of nanoseconds, available in the HLT. In this section, the possibility of using such timing information to improve the trigger efficiency for fractionally-charged particles (FCP) and muons produced in LLP decays is discussed.

In ATLAS, Resistive Plate Chambers (RPC), covering a pseudorapidity range of $|\eta|<1.05$, and Thin Gap Chambers (TGC), covering $|\eta|>1.05$, have single-hit temporal resolutions of about 1.8~ns and 4.0~ns, respectively~\cite{Aaboud_2019_hscp}. Drift tubes, used for precision muon measurements in nearly the full MS, have a hit resolution of 3.2~ns, and Cathode Strip Chambers (CSC), used in the innermost station at $|\eta|>2.0$, have per-segment time resolutions on the order of 3--4~ns~\cite{Aad:2008zzm}. 

In CMS, RPCs have single-hit time resolutions of about 2~ns~\cite{collaboration_2013}. Drift tubes that cover $|\eta|<1.0$ have a single hit time resolution of 2.6~ns, and CSCs, covering $|\eta|>1.0$, have a per-segment time resolution of about 3.4~ns~\cite{CMSMuonDPG}. 

For both ATLAS and CMS, these quoted time resolutions have been obtained offline, after calibrating for differences in the muon time of arrival to the detector and differences in signal propagation due to cable lengths, using prompt muons. It is likely that the picture at HLT will be more complicated.

\subsubsection{Fractionally-charged particles}
\label{s.fcp}

Fractionally-charged particles (FCPs) are considered as a physics case for a delayed muon trigger. Because $dE/dx \sim Q^2$, FCPs often produce signals below threshold in ATLAS and CMS detectors. Standard muon triggers require a well-reconstructed track in both the ID and MS at HLT. The strict hit requirements penalise FCPs, and better efficiency can be obtained with a standalone muon trigger. 

In both ATLAS and CMS, standalone muon triggers in Run 2 used an L1 \pt threshold of approximately 20~\gev, and somewhat higher \pt thresholds at HLT. For example, a 60~\gev HLT threshold was used by ATLAS for most of Run 2, and slightly higher thresholds could likely be envisioned for both ATLAS and CMS for Run-3 conditions. 
For FCPs, the \pt threshold at HLT is not the concern. Previous CMS analyses have required the \pt of the reconstructed FCP candidate to be greater than $\sim 50$~\gev at analysis level in order to reduce backgrounds from $W$ and $Z$ decays~\cite{CMS:2012xi}. Additionally, the reconstructed FCP \pt is larger than the true \pt by the inverse of the particle’s charge. 
Both experiments place stringent hit requirements on the standalone muon track in order to reduce rate from backgrounds. 
On the other hand, the number of hits required on the muon track results in trigger inefficiency for FCPs. FCPs often produce signals below threshold in the drift tubes and CSCs, which operate in proportional mode. Muon trigger chambers, RPCs and TGCs, operate in avalanche or saturated mode, and FPCs are expected to have slightly higher efficiency in those detectors. 

An improved strategy could potentially be to use the same standard L1 muon trigger, and at HLT look for a standalone muon track but with loosened hit requirements, in order to investigate improving efficiency for smaller FCP charges ($Q<\frac{1}{2}$). 
The looser hit requirements at HLT could, in principle, be enabled by using timing data to require that muon hits are delayed, or have a time of flight consistent with $\beta < 1$.
For unexcluded FCP masses~\cite{CMS:2012xi}, the difference in $\beta$ between FCPs and prompt muons is expected to result in delays greater than $\mathcal{O}(1)$~ns.

\subsubsection{Displaced or delayed muons}
\label{sec:dispMuons}

It is not unreasonable to expect that timing information could be useful in searches for LLPs that decay to muons. However, time delays require either slow LLPs or LLPs that decay with large opening angles. In both cases, the limiting factor in trigger efficiency for both ATLAS and CMS is pointing assumptions at L1, which reduce efficiency for muons with transverse impact parameters greater than a few tens of centimeters. 
The CMS Collaboration has performed the successful implementation of a new muon track finder approach that extends the L1 Muon trigger efficiency to cover larger impact parameters, and therefore improves the acceptance for larger lifetimes by as much as a factor of two~\cite{CMS:2019qux}. Such work is also being considered by ATLAS, and is discussed in Section~\ref{s.displacedinMS}. 
However, for this work it is unlikely that adding a delay requirement could be used to further improve efficiency. For displaced di-muon triggers the \pt thresholds at HLT will likely be close to those at L1. We do note that it is important to ensure any implicit timing requirements do not penalise long-lived signatures. 

%%%%%%%%%%%%%%%%%
\subsection{Displaced objects in the muon system at L1}
\label{s.displacedinMS}
%%%%%%%%%%%%%%%%%

\emph{Section Editors: 
Emma Torro, Juliette Alimena}
\vspace*{0.3cm}

\subsubsection{ATLAS}
\label{s.muonatlas}

Neutral LLPs decaying at the end of the HCAL or in the muon system can be triggered in ATLAS~\cite{Aad:2013txa, ATL-DAQ-PUB-2018-002}.
In Run 2 a dedicated HLT trigger selected clusters of at least three (four) muon RoIs in the barrel (end-caps), seeded by a di-muon L1 trigger. 
Reducing the minimum number of RoIs required in the cluster could increase the signal efficiency, specially for low-boosted LLPs. 
Studies are needed to understand what the gain in signal efficiency would be and how to deal with background reduction if this condition is relaxed.

ATLAS can also trigger on displaced muons at HLT using the L1 seeds for isolated activity in the MS. Since pointing assumptions at L1 are a major limiting factor for the efficiency of L1 triggers for displaced muons, the reach for impact parameters larger than tens of centimeters could be significantly improved with a dedicated L1 trigger.

\subsubsection{CMS}
\label{s.muoncms}

Although no L1 dedicated triggers were implemented to select displaced leptons of hadrons in the CMS experiment during the Run 2, improvements to barrel muon track finding, referred to in Section \ref{sec:dispMuons}, were developed to include this capability. The successful commissioning of the kBMTF algorithm~\cite{CMS:2019qux} using cosmic data demonstrated the performance of this algorithm in selecting displaced objects~\cite{CERN-LHCC-2020-004}. The Run-3 menu is integrating these new features to expand the physics acceptance of the CMS detector.

Lepton triggers at L1 in the muon system (MS) of CMS for Run 2 relied upon a vertex constraint, making them not particularly sensitive to displaced muons. The standard methods of triggering on muons in both the barrel and end-caps at L1, the Barrel Muon Track Finder and the End-cap Muon Track Finder~\cite{CMS:2020cmk}, are being improved using advances in track-finding (such as improved Kalman filter approaches and machine learning techniques) to trigger on displaced muons without losing sensitivity to those from prompt decays~\cite{CMS_L1_displaced_barrel,CMS_L1_displaced_endcap,CMS_Trigger_Acosta,CERN-LHCC-2020-004}. For example, methods being developed for the Phase 2 upgrade show great promise for improving the efficiency of triggering on displaced muons in the end-caps~\cite{CMS-TDR-013,CMS-TDR-016}, and it would be highly beneficial to investigate such methods for use in Run 3 as well.

Besides decays to displaced muons, the CMS Collaboration is exploring the ability to trigger on hadronic decays in the muon system. For example, neutral LLPs, such as sterile neutrinos, can decay and produce a shower of particles in the muon system end-caps. These showers could result in a large number of track segments and hits in a single muon system chamber, which should in principle be able to be triggered upon. Furthermore, the Cathode Strip Chamber L1 trigger is currently being upgraded during the Long Shutdown 2 with new hardware and firmware~\cite{CMS_Muon_L1_Talk}, providing additional usable bandwidth to identify these high-multiplicity events.

In addition to these efforts, further studies are underway to ensure CMS does not miss any opportunity to trigger decays in the MS at L1 during Run 3.

%%%%%%%%%%%%%%%%%
\section{LHCb}
\label{s.LHCb}
%%%%%%%%%%%%%%%%%
\emph{Section Editors: 
Carlos Vázquez Sierra, Federico Leo Redi, Lorenzo Sestini, Phil Ilten, Xabier Cid Vidal, Murilo Santana Rangel, Maarten van Veghel, Adrián Casais Vidal, Mike Williams, Arantza Oyanguren, Brij Jashal, Louis Henry, Giovanni Punzi}
\vspace*{0.3cm}

The LHCb experiment at CERN, originally designed for the study of decays of heavy flavour particles and instrumented in the forward region (pseudorapidity between 2 and 5), has proven to also serve as an excellent general-purpose detector for beyond SM (BSM) LLP searches in the past years. LHCb is especially capable to access low masses and low lifetimes, which are regions that are very difficult to access for other LHC experiments like ATLAS and CMS, and where LHCb has unique and complementary capabilities.
This is possible thanks to the outstanding vertex reconstruction capabilities of the LHCb Vertex Locator (VELO) (allowing to resolve very short lifetimes of around few $ps$), to the excellent track momentum resolution of the detector (especially for tracks that traverse the full LHCb tracking system and originate from particles decaying within the VELO region, allowing to perform an exclusive reconstruction of $B$-mesons and $D$-mesons), to the particle identification capabilities provided by the electromagnetic (ECAL) and hadronic (HCAL) calorimeters, the muon system, and the Ring Imaging Cherenkov (RICH) sub-stations (allowing to distinguish between light hadrons, such $K$ and $\pi$), and to the LHCb trigger system \cite{Alves:2008zz}. 

The LHCb detector is undergoing through to a major upgrade for Run 3, in particular, a better performance is expected from all the sub-systems previously mentioned \cite{Collaboration:1624074, Collaboration:1624070, Collaboration:1647400}. In particular, the LHCb trigger system consisted, during Run 1 and 2, of a Level-0 (L0) hardware stage, and two High-Level Trigger software stages, namely, HLT1 and HLT2. 

For Run 3, the upgraded LHCb trigger system \cite{CERN-LHCC-2014-016} will feature the removal of the L0 stage (allowing to access very soft transverse momentum states, of a few \gev/\mev for heavy flavour jets/light hadrons) and real-time reconstruction and selection at HLT, where the selective persistency model used for Run 2 will become the baseline for Run 3 as well \cite{Aaij:2019uij}. 
The HLT1 will be entirely implemented on GPUs with great networking cost reduction associated with sending data to the event filter farm~\cite{Aaij:2019zbu, LHCbCollaboration:2717938}. A comparison of performance and implementation details between a CPU-based and a GPU-based approach for the HLT1 is summarised in Ref.~\cite{Aaij:2021mzf}. The cost savings can be used to improve the trigger throughput and possible scenarios with a more flexible and/or more dedicated requirements at HLT1 can be discussed.
Since most of the LHCb physics programme is based on muons and/or displaced tracks in final states, and due to the necessary soft requirements on these tracks, a global event cut to remove the 10\% busiest events is planned to be used. The typical minimum transverse momentum requirement for a single displaced track trigger is 1~\gev. If additional event requirements are applied, the global event cut may be removed and the minimum transverse momentum requirement reduced. 

Much of the LLP sensitivity already demonstrated by LHCb in Run~2 and the improvements expected in Run~3 are serendipitous: LLP physics was not considered when designing the current or soon-to-be upgraded LHCb detector, but since studying LLPs often benefits from the same experimental capabilities required to study heavy flavour decays, LHCb has excellent performance in this area. 
However, it is worth noting that for many types of BSM physics, especially for bosons that couple to a non-conserved current, $b$-hadron penguin decays will be the dominant production mode for masses less than $m(B)-m(K)$~\cite{Dror:2017ehi}.
In such cases, which includes \gev-scale axion-like particles and scalars that couple to the SM via the Higgs portal~\cite{LHCb-PAPER-2015-036,LHCb-PAPER-2016-052}, LHCb has been purpose built to provide world-leading performance.
Potential ideas for Run 3 (beyond 2021) which would benefit from this upgraded trigger system, are presented in the following sub-sections. Most of these ideas have been also discussed in a more general way in Ref.~\cite{Borsato:2021aum}, whereas the contents of this section will focus on how to trigger these signatures with the upgraded LHCb experiment. 

%%%%%%%%%%%%%%%%%%%%%%%%%%%%%%%%%%%%%%%%%%%%%%
\subsection{Displaced jets}
\label{s.lhcbdisplacedjets}

We discuss two new triggers at HLT2 for LHCb searches targeting LLPs decaying to jets (this is to be distinguished from triggers for exclusive hadronic LLP decay modes in the next section):
\begin{enumerate}
    \item  A ``displaced fat jet'' trigger, using tracker and calorimeter information to construct the fat jet containing displaced tracks (displaced vertices are not part of the reconstruction, similar to the CMS displaced jet trigger). The aim is to probe LLP masses below $\sim 25~\gev$, where existing displaced di-jet triggers lose sensitivity.
    \item A trigger that reconstructs displaced tracks of some minimum multiplicity/impact parameter into one or multiple displaced vertices. This is already done in off-line analyses, but performing this reconstruction at trigger level could greatly increase acceptance. Such a technique could be very flexible and inclusive with regards to LLP mass. 
\end{enumerate}

The first idea represents a natural evolution of existing searches at for LLPs decaying into a pair of heavy-flavour jets, in the di-jet invariant mass range between 25 and 50~\gev, using Run-1 data~\cite{Aaij:2017mic}. %
Such a signal is produced in many models, including LLP production in exotic Higgs decays.
In this search, a pair of $b$($c$)-jets with \pt greater than 5~\gev, associated to a displaced secondary vertex has been required. At low masses, the two jets from the LLP decays could be collimated and they may merge together producing a single fat jet \cite{Plehn:2009rk}.

The reconstruction and identification of displaced fat jets in the HLT2 may be the ideal strategy to push the lower mass limit of the searches down to 5--10~\gev. 
First of all, fat jets may be clustered using the anti-$k_t$ algorithm, using displaced tracks and calorimeter clusters as inputs, selected by a particle-flow algorithm. 
The fat-jet radius parameter should be optimised using simulation, and it is usually chosen in the range between 0.8 and 2. 
In Run 2, LHCb demonstrated its capability in reconstructing jets at HLT2 level with an optimal time performance.
The measurement of the fat-jet mass can be improved by applying a grooming technique like MassDrop Tagger~\cite{Marzani:2017mva} or SoftDrop~\cite{Larkoski:2014wba}.
As a second step, several techniques can be used to identify the fat jets and to distinguish them from the QCD single-jets background.
Moreover, $b\bar{b}$, $c\bar{c}$ or light fat jets (produced by the fragmentation of a pair $u$, $d$, $s$ quarks or gluons) could be separated, improving the sensitivity in specific decay channels.

Similar techniques to those employed by ATLAS and CMS in the identification of boosted objects by exploiting the jet-substructure~\cite{Khachatryan:2014vla, Aaboud:2018psm, Sirunyan:2020hwz} could be used for such classification.
These techniques are based on ML algorithms, using classifiers such as Deep Neural Networks~\cite{jet_substructure}. Several other options can be explored, like Graph Networks \cite{Ju:2020tbo}, Generative Adversarial Networks \cite{Carrazza:2019cnt}, Particle Clouds \cite{Qu:2019gqs} or Tree Tensor Networks \cite{Trenti:2020ceh}, that have been already used with success in jet physics.
Two different kind of inputs, or a combination of the two, could be used in this task. The first type of input is the fat jet image, defined as the distribution of the measured energy in the $(\eta, \phi)$ space, arranged in a two-dimensional matrix. 
The second type of input is the list of observables related to fat jet constituents, which are ordered by decreasing impact parameters or decreasing transverse energy for some of the listed algorithms. 
The application of ML algorithms could be implemented at HLT2 level, and a cut on the classifier output could be applied to select the signal and remove the background. Since the time performance of the application is in general dependent on the classifier complexity and on the number of input observables, those have to be optimised in order to have an adequate time performance for the HLT2. 

In Run 3 a higher pile-up with respect to Run 1 and 2 is to be expected. This may have an impact on the jet reconstruction performance, in particular it can degrade the jet energy resolution and increase the fake-jet rate. Techniques could be implemented in order to subtract the pile-up at jet-constituents level, as it is done with the SoftKiller algorithm~\cite{Cacciari:2014gra}. Other tools based on ML could be employed, like the Pile-up Mitigation with ML (PUMML)~\cite{Komiske:2017ubm}. The usage of such techniques at in Run 3 could improve the sensitivity in the search for LLPs.

During Run 1 and 2, the LHCb level-0 trigger applied a global event cut (GEC) on the hit multiplicity in the scintillating-pad detectors at hardware level. Di-jet events have high hit multiplicities, therefore this requirement accepted only 60\% of di-jet events in Run 2. Moreover the GEC introduce a systematic uncertainty in the efficiency determination, since at LHCb the simulated hit-multiplicity distribution shows discrepancies with data. In Run 3 the hardware trigger will be removed and a full software trigger will be available. For this reason the GEC can be removed from the LLP trigger lines, improving the selection efficiency.

The second idea relies on detailed reconstruction of one or more displaced vertices within the high-multiplicity final state of the LLP decay. 
In the Run-1 search for displaced di-jets mentioned above, triggers for displaced vertices have been used: pre-selected tracks with a minimum IP requirement are used to reconstruct a displaced vertex, required to be highly displaced (more than 0.4 mm in the transverse plane), to have a minimum track multiplicity, and to have an invariant mass higher than the typical $B$-meson masses.
However, a new approach has been considered for Run-2 searches, consisting in the development on new triggers where displaced tracks are selected, and then used to reconstruct a displaced jet with a cone radius of ${\Delta}R<0.5$, employing the anti-$k_t$ clustering algorithm. After the displaced jet is reconstructed, very loose minimum displacement and minimum \pt requirements are imposed. A vertex reconstruction algorithm is also ran using the jet tracks as input tracks, and reconstructing all displaced vertices inside the jet cone. Finally, requirements on the number of both displaced vertices and on the daughters of the displaced vertices are imposed: these numbers are chosen depending on the mass of the jet and its flavour. For $b$-jets, when the jet invariant mass is reasonably above the $b\bar{b}$ threshold, the number of displaced vertices can be as high as three or four, requiring at least five tracks per displaced vertex (which corresponds to the average track multiplicity of the decay of a $B$-meson). For $c$-jets, numbers are different and have to be properly tuned taking into account the mass of the jet and the average track multiplicity of the decay of a $D$-meson. 

This new technique is extremely flexible and allows to trigger on signatures of displaced heavy-flavour jets, without imposing any requirement on the jet invariant mass and requiring a loose threshold cut for track displacement. Furthermore, a multiplicity of these displaced jets can be required as well to reduce background as much as possible (especially for very low track displacements). A first version of these triggers has been included in the 2018 data-taking campaign, and proven to be efficient. However, even if it looks promising enough, for Run 3 it would be required to improve the selection used in these triggers by performing dedicated optimisation studies. 

A potential new idea which could be worth investigating, even if not directly related to BSM LLPs, would be to also try to use these technique to perform jet flavour tagging. 

%%%%%%%%%%%%%%%%%%%%%%%%%%%%%%%%%%%%%%%%%%%%%%
\subsection{Displaced light hadrons}
\label{s.displacedhadronslhcb}

A great advantage of LHCb compared to the main detectors is the ability to study exclusive hadronic decays in great detail. This capability can be put to use to search for low-mass hadronically decaying LLPs that occur in many models (see {\it{e.g.}}\ Ref.~\cite{Beacham:2019nyx} and discussions therein) and are difficult to search for at ATLAS/CMS. We discuss two concrete HLT2 trigger ideas for LHCb to increase reach for these scenarios: 
\begin{enumerate}
    \item Complementary topological triggers for displaced hadronic decays, where kinematic requirements should be tuned in a way that displaced objects with masses and lifetimes below those values typical of a $B$-meson can be selected. 
    \item Dedicated triggers for exclusive hadronic decay modes like LLP $\to$ $K^+ K^-$, $\pi^+ \pi^-$, \ldots that are highly motivated for LLPs with masses of $\mathcal{O}(1~\gev)$ in specific simplified models like singlet scalar extensions. Requiring multiple displaced vertices in the event would suppress backgrounds while retaining high efficiency for short lifetimes.
\end{enumerate}

Since Run 1, LHCb has benefited from the so-called {\it{topological triggers}}~\cite{BBDT, Likhomanenko:2015aba} at the HLT2 software-based stage, used to trigger in an efficient way any decay of a $B$-meson with at least two daughters. These triggers are based on an iterative method with two to one object combinations, where pre-selected tracks ($\pt>0.5$~\gev and $p>5$~\gev, as well as track and IP $\chi^2$ requirements to remove ghosts and prompt backgrounds) are combined to form a two-body object, then, an additional track is combined as well to form a three-body object, and so on. The shortest distance between two objects in each step should be less than 0.15~mm. Further requirements on the \pt of the children of these objects are applied (hardest daughter should have $\pt>1.5$~\gev), as well as minimum invariant mass above 2.5~\gev and IP $\chi^2$ requirements on the objects themselves are considered, in order to remove background from $D$-mesons. At least one of these multi-body objects is required to be present in the event. 
These triggers were flexible enough to select LLPs produced in $b$-hadron penguin decays in Runs 1 and 2, and we hope they will continue to do so in Run~3. 
In addition, exclusive penguin decays should be targeted that are known to be highly sensitive to high-profile LLP models, {\em e.g.}, ALPs that couple to gluons~\cite{Aloni:2018vki}.

However, despite the fact these triggers have been designed to be as much inclusive as possible, there are requirements on the mass and the flight distance of the multi-body object which corresponds to those of a $B$-meson, effectively reducing the reach of these triggers for lower lifetimes and masses. A possible solution would be to tune these requirements to extend, at least, the lifetime reach to a reasonable threshold value, which would be beneficial for models that predicts scenarios where an LLP decays hadronically within the detector acceptance. However, further studies would be needed since relaxing or just fully removing these requirements would probably lead to a critical increase in the trigger input rate, due to the presence of prompt and moderately displaced hadronic QCD decays. 

Another potentially very powerful approach is to develop dedicated triggers for multiple displaced vertices in the event, decaying into two pairs of light hadrons ($K^+K^-$, $\pi^+\pi^-$, or even $K^\pm\pi^\mp$)~\cite{CidVidal:2019urm}. This would target decay modes that are predicted {\it{e.g.}}\ for light scalar LLPs that decay through the Higgs portal. These triggers would use pre-selected tracks identified as light hadrons (by requiring a reasonable good likelihood response from the RICH detectors), where a minimum impact parameter and loose kinematic requirements (minimum-\pt requirement of 2~\gev) would be imposed, to veto light hadrons coming from the PV. 
Furthermore, good vertexing would be required for the reconstructed di-hadron pair, as well as to be pointing to the PV. However, these requirements would not be sufficient in order to have a reasonable input rate: there would be a significant presence of backgrounds from hadronic $K_S^0$ decays and $D^0$ decays (either produced in the PV or from the decay of a $B$-meson), as well as from $\Lambda$-baryon decays into a proton and a light hadron (where the proton is mis-identified as a $K$, or even less likely as a $\pi$ by the RICH system).

To remove these backgrounds, one possibility would be to have more strict particle identification requirements on the hadron tracks and on the di-hadron pair displacement, which would reduce the reach of these triggers for low lifetimes, depending on how strict the requirement on the displacement is chosen to be. 
For sufficiently long lifetimes, contributions from these backgrounds would be highly suppressed, which might make possible to develop a single displaced vertex trigger selection while keeping the input rates under control: however, this would need to be studied in very detail. 

Another possibility would be to impose a requirement on the di-hadron pair multiplicity in the event, which would lead to a very effective background removal, but also to make these triggers less useful: only models which predict a multiplicity of soft displaced pairs of light hadrons in the final state and in the LHCb acceptance would benefit from those~\cite{Pierce:2017taw} (this is not the ideal case for, {\it{e.g.}}\ models where two di-hadron pairs are produced directly from an exotic Higgs-boson decay, since the \pt of the pairs will be large enough to have, most of the time, only one of the pairs decaying inside the LHCb acceptance). 
For the 2018 data-taking campaign, offline selections with these multiplicity requirements have been implemented, and proved indeed to be highly efficient with small multiplicities (at least two di-hadron pairs in the event). 

There is another interesting possibility, where certain models predict the production of pair of meson resonances (like $K^*K^*$) via a light scalar LLP as well. For this particular case, each meson would decay into two or three hadrons, leading to a displaced vertex with more than two light hadron tracks. A similar strategy described in the previous paragraphs could be considered for this case, however, a re-optimisation of the selection is mandatory ({\it e.g.} requirements on the impact parameter should be relaxed to account for more than two tracks in the decay vertex). 

In general, it is expected that the fully hadronic final states will have a higher trigger efficiency in comparison with Run 2, specially if dedicated triggers for low mass displaced vertices are developed. 

In the regard of HLT1, a baseline approach for Run 3 is to use a VELO-only Kalman filter in the sequence, if the impact parameter is considered more important than the momentum measurement. This would lead to a significant computing speed-up compared to the application of a full Kalman filter, used only for fake track rejection at HLT1. 

%%%%%%%%%%%%%%%%%%%%%%%%%%%%%%%%%%%%%%%%%%%%%%
\subsection{Displaced (di-)muons and (di-)tau leptons}
\label{s.displacedditaulhcb}

LHCb has a very efficient set of triggers for displaced di-muon searches, consisting of kinematic and particle identification requirements for muon tracks, as well as an invariant mass and vertex displacement requirement. These triggers have been widely used in LHCb, {\it{e.g.}}\ to select $Z^0\to\mu\mu$ (without the vertex displacement requirement)~\cite{Aaij:2016mgv} or $B_s^0\to\mu\mu$~\cite{CMS:2014xfa} decays, as well as for the search of new prompt di-muon resonances~\cite{Aaij:2019bvg, Aaij:2020ikh, Aaij:2018xpt}. There is little room for improvement for these kind of signatures (beyond that removing the hardware-trigger stage will lead to a large increase in muon efficiencies at low \pt), other than developing dedicated triggers where mass and vertex displacement requirements are tuned accordingly to extend the reach of the triggers. 
We therefore focus on final states with one or more displaced tau leptons, and discuss two new types of HLT2 triggers:
\begin{enumerate}
    \item A dedicated trigger for displaced $\tau\tau$, arising {\it{e.g.}}\ in LLPs decaying through the Higgs portal. This would be particularly motivated for LLP masses below $\sim 50~\gev$ (even up to $\sim 100~\gev$) where LHCb can be highly competitive with ATLAS/CMS. 

    \item A trigger for events with a single prompt and a single displaced tau that would be sensitive to right-handed neutrino-type models. 

\end{enumerate}

A variety of searches involving displaced tau leptons have been done by LHCb: either produced in the semi-leptonic decay of $B$-mesons and decaying hadronically (three-prong) (where {\it{topological triggers}} are used)~\cite{Aaij:2017deq}, or simply pair-produced from the purely leptonic decay of a $B_s^0$ meson~\cite{Aaij:2017xqt}, among others. The latter is an analysis which uses very specific data-driven methods and special vertex requirements, where both tau leptons are reconstructed only from their hadronic decays, selected with the {\it{topological triggers}} previously described. 
Furthermore, several studies have been done in the regard of trigger efficiencies for Run-3 conditions. In particular, the trigger efficiency of the {\it{topological triggers}} for three-prong $\tau$ lepton decays it set to increase from 17\% in Run 1 to 64\% in Run 3~\cite{CERN-LHCC-2014-016}.

New dedicated triggers could be developed for displaced vertices decaying into pair of $\tau$ leptons, where the pair points to the PV. These triggers could be used for searches of LLPs decaying into $\tau\tau$. 
This is most motivated for LLP masses below $\sim 50~\gev$ where LHCb could be highly competitive with the ATLAS and CMS detectors. 
A similar strategy to that followed by LHCb in the search for $Z^0\to\tau\tau$~\cite{Aaij:2624023} can be followed: exploit different combinations of the two $\tau$ lepton decay modes, namely, $\tau_1\tau_2$ where (a) $\tau_1\to\pi\pi\pi\nu$ and $\tau_2\to{l}\nu\nu$, or where (b) $\tau_1\to{l_1}\nu\nu$ and $\tau_2\to{l_2}\nu\nu$ having $l_1$ and $l_2$ different flavours. Unfortunately, both of these options have their own benefits and flaws. 

On the one hand, (b) will consist on the combination of two different-flavour lepton tracks, pointing to the same decay vertex (with a displacement compatible with the $\tau$ lifetime), and well isolated (with respect to other tracks in the event). This will effectively help to remove backgrounds from leptonic decays of SM particles. However, given the fact that LHCb is not a hermetic spectrometer, none of the tau leptons can be reconstructed, since there are two neutrinos per $\tau$ lepton in the final state. No information on missing particles makes the removal of combinatorial background almost impossible. 

On the other hand, (a) would consist of an almost well-triggered $\tau_1$ (where only one missing neutrino would affect the vertex reconstruction), and an isolated lepton, all pointing to the same displaced decay vertex. This will indeed lead to a penalty on the reconstruction efficiency since there are four tracks in the final state, plus the fact that the hadronic $\tau$ decay has a smaller branching fraction that the purely leptonic mode, but it will remove a large component of pions from the QCD background. A third possibility, albeit much less efficient, would be to consider both $\tau_1$ and $\tau_2$ to decay hadronically, effectively removing a very large component of the QCD background, but reducing the reconstruction efficiency in a substantial way due to the presence of six tracks in the final state. 

Heavy neutral lepton LLPs are produced in association with a prompt lepton and have a displaced decay that includes a charged lepton of generally the same flavour in the final state. 
This motivates development of a trigger for a prompt $\tau_1$ + displaced $\tau_2$ (with efficiency for muons already being near-optimal for the reasons discussed above). %
Background rejection generally requires the displaced $\tau$ to decay hadronically. 
However, for the prompt $\tau$, a leptonic decay would have the least background, but it may also be possible to include hadronic prompt $\tau$ in the search. Given the fact that the direction of flight of the prompt $\tau$ lepton can be known if assumed to be approximately originated at the proton-proton interaction point, it can be efficiently triggered by imposing a requirement in the so-called {\it{corrected mass}}~\cite{Abe:1997sb}, used in many LHCb publications involving $\tau$ leptons in the final state: in the prompt $\tau$ direction of flight (which should be known as already stated), and if assuming the neutrino to be massless, the $\pi\pi\pi$ system and the neutrino should have opposite $\pt$, hence, the {\it{corrected mass}} of $\tau_1$ would be $\sqrt{m^2(\pi\pi\pi) + \pt^2(\pi\pi\pi)} + \pt(\pi\pi\pi)$ (all of these quantities should be projected in the prompt $\tau$ direction of flight), which becomes a good proxy to the mass of the $\tau$ lepton, and can be used as a discriminant variable. 
Therefore, triggers for prompt leptonic (or hadronic) $\tau_1$ and a displaced hadronic $\tau_2$ should be closely investigated.

%%%%%%%%%%%%%%%%%%%%%%%%%%%%%%%%%%%%%%%%%%%%%%%%
\subsection{Displaced (di-)electrons and (di-)photons}
\label{s.displaceddiphotonselectronslhcb}

In Run 3, L0 hardware trigger requirements will be removed. This allows the deployment of two new triggers targeting low-mass LLPs: 
\begin{enumerate}
    \item displaced di-electrons for LLP masses below $\sim 100~\gev$, and
    \item displaced di-photons for LLP masses below $\sim 20~\gev$  (due to ECAL energy saturation if using unconverted photons). 
\end{enumerate}
This will greatly expand our physics reach for BSM scenarios like dark photons, axions and dark showers. 
Below, we first review the L0 limitation of Run 1 and 2 before discussing how these new triggers could be implemented.

In Run 1 and 2, selections on final states with di-electrons and di-photons were limited by stringent requirements at L0 level on deposits in the ECAL. A good example of what was possible at that time is the following: for Run 2, di-photon selections were first introduced in 2015 at LHCb for the search of the rare decay $B_s^0 \to \gamma \gamma$. The reconstruction of photons in LHCb can be classified in two big categories, detailed as follows:  converted photons, when photons undergo a conversion into a $e^+ e^-$ pair upon interaction with the material budget, provided the tracks and the secondary vertex are successfully reconstructed; and unconverted photons, where the photons do not interact until they reach the ECAL, where they are reconstructed using the energy deposit on the corresponding cluster. Spatial and energy resolution is better for converted photons, but only a fraction of these will undergo a conversion. Hence, unconverted photons play a significant role, for instance, on analyses statistically limited. 
The use of combinations of ECAL clusters from L0 was introduced in order to reconstruct unconverted photon candidates at the HLT1, since no calorimeter reconstruction -- not even partial -- was present at this stage of the trigger sequence. Reconstruction of unconverted photons, based on L0 clusters, has the drawback of not being possible to effectively impose a requirement on $E_T$ values over 10~\gev, due to saturation of these ECAL objects. Dedicated reconstruction during Run 3 will allow for more efficient requirements on photons with high $E_T$ \cite{Collaboration:1624074}. 

In addition, searches for di-photons with conversions benefit from the selections already available for displaced tracks, which are designed to select di-electron tracks from conversions. Additionally, a trigger based on a Neural Network classifier has been introduced during the 2018 data-taking campaign, in order to correctly identify pairs of photons (both relying on conversions and not). This was the first time such kind of classifiers were used to select this type of multi-body decays \cite{CidCasaisPhotons:2019}. A similar approach could be followed during Run 3, designing a similar classifier including features to account for the higher pile-up and multiplicity in the event. 
An inclusive di-electron trigger was also implemented during Run 2, which again used electromagnetic clusters from the L0, but now linked against tracks. Triggers similar to the inclusive di-muon lines used to search for low masses resonances, with \pt thresholds as low as $1~\gev$, were introduced. Additional inclusive displaced lines, as well as exclusive $\pi^0 \to e^+ e^- \gamma$ and $\eta \to e^+ e^- \gamma$ triggers were also implemented. While no analyses have yet been published using these lines, initial studies indicate the efficacy of these triggers. 

For Run 3, with the removal of the L0 level, a direct selection based on information from the tracking system is possible, both at HLT1 and HLT2, allowing to select on more discriminate information like displacement, vertex quality and others, in comparison to just ECAL information. In addition to the inclusive triggers using tracks, the ECAL information will also be decoded and available in the HLT1 selection. The information obtained from the full ECAL reconstruction can in turn be matched with the information from the tracking systems, opening up the identification of electrons within the set of available charged tracks at HLT1, which was only possible during Run 2 using the more rudimentary L0 electromagnetic clusters. In addition, the use of partial track-reconstruction for di-electrons can considerably decrease material-interaction related inefficiencies, which can be implemented at both HLT1 and HLT2. With this new features in hand, dedicated lines can be developed for displaced electron searches.

Neutral long-lived particles can be searched by using ECAL clusters that are isolated from tracks and clustered by a simple jet reconstruction. Finally, neutral hadrons appear in the dominant decay modes of models like ALPs that couple to gluons, and in Run~3 it will be possible to search for decays like $a \to \pi^+\pi^-\pi^0$ in HLT1. 

The aforementioned additions to the trigger system of LHCb open up the selection of very soft (di-)electrons and converted (di-)photons that are predicted BSM scenarios like dark photons, axions, and dark showers. 
A SM target of an inclusive low-mass displaced di-photon resonance search is the search for  muonium~\cite{CidVidal:2019qub} using $\eta \to \gamma \textrm{M}(\to e^+ e^-)$, which requires both low \pt electrons and photons. 
With the full clustering of the ECAL, bremsstrahlung recovery will also be available at the HLT1, and will help increase momentum resolution and improve sensitivity in resonance searches. For exclusive final states with photons and electrons, this can significantly reduce trigger rates.

%%%%%%%%%%%%%%%%%%%%%%%%%%%%%%%%%%%%%%%%%%%%%%%%%%
\subsection{Track signatures beyond the inner detector}
\label{LHCb.downstream}

\begin{figure}
    \centering
    \includegraphics[width = 0.55\textwidth]{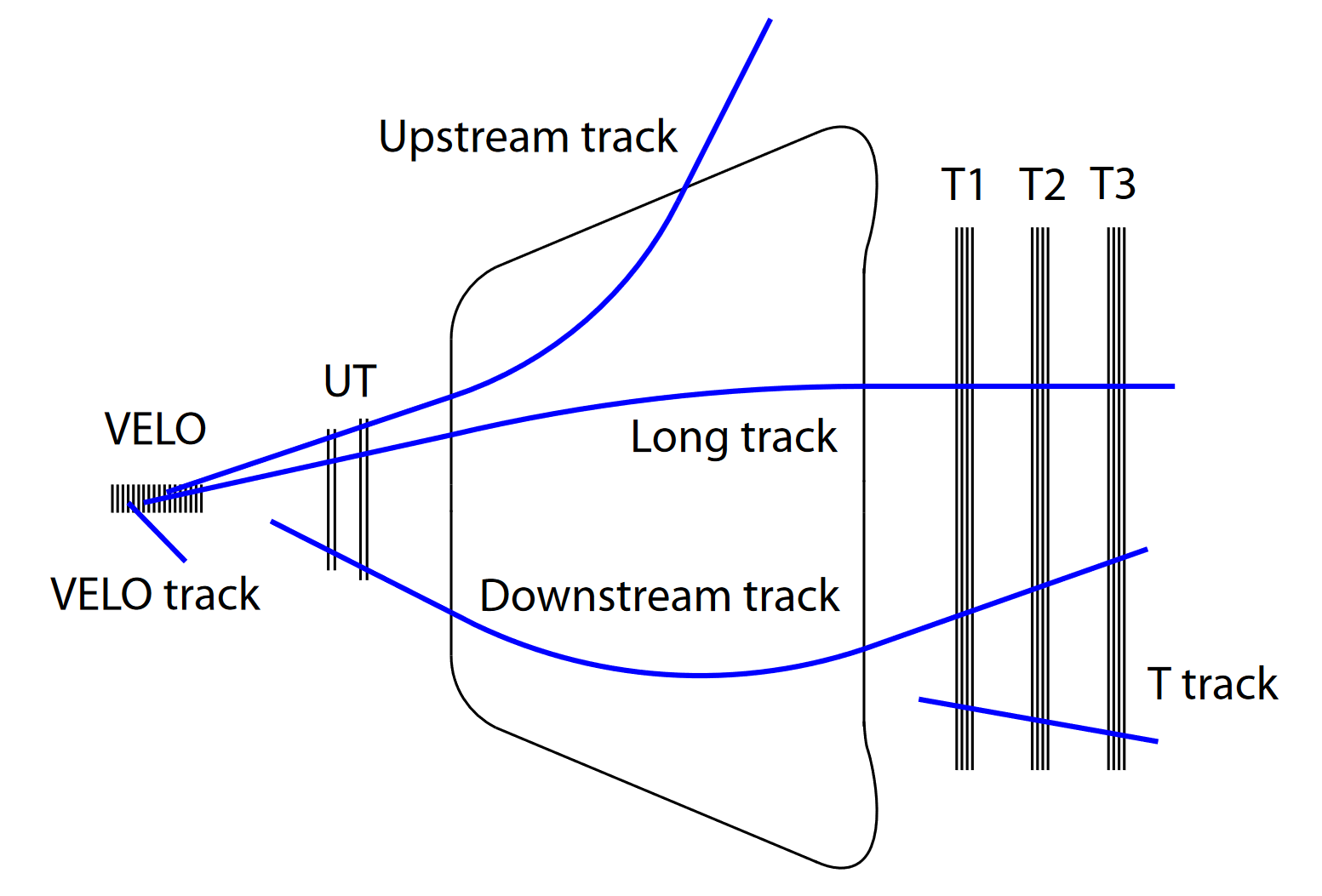}
    \caption{Track types reconstructed at the LHCb experiment by the tracking system~\cite{Collaboration:1647400}.}
    \label{fig:LHCb_tracks}
\end{figure}

Most of the LLP searches are based on displaced vertices and tracks in the inner detectors, or on signatures in the muon and calorimeter systems. If an unstable neutral LLP with long lifetime ($\tau$\,$\sim$\,ns) decays to SM-like charged tracks after the first tracker (VELO) and before reaching the calorimeter system, the present LHCb HLT1 trigger system will not be able to reconstruct and select them. This is due to the fact that the current strategies for LLP searches at LHCb rely on the so-called {\it{long}} tracks, this is, tracks reconstructed with information from all the tracking sub-stations, including hits in the VELO system (which is limited to a radial displacement between 0 and 30 mm and to a transverse displacement between 0 a 500 mm, in cylindrical coordinates). Figure \ref{fig:LHCb_tracks} shows an sketch of the track types reconstructed at LHCb. If no other long track is produced in addition to the LLP, the expected trigger efficiencies are below few $\%$, letting away the possibility to perform any kind of search. 

However, the reconstruction of such {\it{downstream}} tracks (where information from hits in all tracking sub-stations except in the VELO are used), is an interesting possibility to consider for the upgraded HLT1 system, making use of information from the Upstream (UT) and Scintillating Fibre (SciFi) trackers \cite{Collaboration:1647400}. These tracks are typically associated to vertices transversally displaced more than 500 mm (outside the VELO acceptance), allowing to reach further vertex displacements and higher LLP lifetimes. 

There are a lot of efforts at present to implement the downstream tracking at the HLT1 level, using information from hits in the tracking sub-stations except the VELO. The main issue of such trigger implementation is the high amount of combinatorics and the timing requirement to make a decision. 
Nevertheless with the new heterogeneous computing approach, and the increase in the trigger throughput this is expected to be feasible. 
LHCb is already introducing GPUs for its HLT1 in the upcoming Run 3.
For future runs, the potential of a specialised FPGA-based tracking device operating before HLT1 has been in consideration since the LHCb's EOI for future upgrades \cite{Aaij:2244311}. Further R\&D has been carried out since then \cite{CENCI2019344,LHCbReal-TimeAnalysisproject:2020jik,Cenci:2020fbz}, and the collaboration is now including a dedicated FPGA-based Downstream tracker (DWT) in its list of possible enhancement for Run 4.
HLT2 topological lines with two or more tracks could be adapted to the case of downstream tracks, with relatively similar track and vertex quality requirements and without any mass constraint. Dedicated ML tools can be developed and implemented here to reduce the ghost tracks rate. 
As a step towards this goal, a HLT1 trigger for $K^0_S$ and $\Lambda^0$ particles (limited to decays within the VELO volume) is currently being commissioned for deployment in Run 3. This is likely going to be a useful pilot for the development of more general-purpose LLP lines at HLT1. 
A downstream track trigger would add significant and very general capability to search for many kinds of LLP decays with decay lengths greater than $\mathcal{O}(10~cm)$, for both leptonic and hadronic decay modes.
BSM targets include exotic Higgs decays to LLPs, as well as 
charginos and sleptons searches in some SUSY scenarios decaying into final states with two muons with a lifetime range of 0.1--10~ns and large masses. The information of the two muons could be combined with calorimetric and other track signatures. The parameter phase space can also be widened for Higgs-like searches in mSUGRA models with R parity and baryon number violation, where the Higgs decays into two neutralino particles \cite{Kaplan_2007}. Dark photon searches could also benefit for small off-shell photon mixing factors. Additionally, LLPs decaying into hyperons or associated to them will automatically be possible.

%%%%%%%%%%%%%%%%%%%%%%%%%%%%%%%%%%%%%%%%%%%%%%%%%%
\subsection{Other challenging ideas}
\label{s.lhcbchallengingideas}

A number of ideas for potential exotic searches in LHCb, which would make use of brand new techniques yet to be developed, have been discussed over the past months:
\begin{enumerate}

\item 
A search for fractional charged objects, making use of the RICH sub-detectors for particle identification, would be a first idea. The number of photons produced by the Cherenkov effect and measured by the PMT is proportional to $Q^2$ of the particle transversing the RICH station, hence, a fractional charged object with a charge of {\it{e.g.}}\ $e/3$ would produce a fraction of $1/9$ photons with respect to a SM charged particle with charge $e$. This would lead to a significantly different response from the RICH sub-detector, which could be use as a powerful discriminant for this analysis. However, this kind of searches would need dedicated simulation studies (which has never been done before in LHCb), as well as dedicated triggers based on this discriminant requirement. 

\item 
LHCb has published a proof-of-concept analysis of a search for Charged Massive Stable Particles (CMSP) \cite{Aaij:2015ica}. Again, this search relies on the response from the RICH sub-detectors: a CMSP would be a very slow ionising particle passing through the RICH stations. Since the cosine of the opening angle of Cherenkov photons is inversely proportional to the velocity of the particle, a CMSP would produce Cherenkov photons highly collimated in the direction of the particle, well below the sensitive threshold of sub-detector, leading to a charged track transversing the LHCb detector with an absence of signal from the RICH system. This is a very characteristic signature, highly discriminant, which would lead to a zero-background search. Unfortunately, LHCb capabilities in this proof-of-concept turned out to not be competitive with existing upper limits, however, it would be worth to consider revisiting this technique again and searching for this signature in the coming years, since a brand new RICH system will be installed in LHCb for Run 3. 

\item
Another idea would be to consider potential triggers on SUEPs (soft, unclustered energy patterns) \cite{Knapen:2016hky}, where the cluster occupancies in both the ECAL and HCAL could be used as discriminant variables. However, dedicated simulations studies would be needed to discuss the feasibility of such a trigger strategy in LHCb. 

\item 
Some very specific algorithms which rely only on track information (available at HLT1) might be studied as well, as {\it{e.g.}} computing {\it{event isotropy}} as described in \cite{Cesarotti:2020uod} could help tag SUEP events, and using this information as a discriminant variable for a potential trigger selection. 

\end{enumerate}
 Most of these ideas need dedicated preliminary studies, mainly focused on the development of dedicated simulation and reconstruction algorithms. Because of this, the preparation and commissioning period for Run 3 turns out to be the ideal time-window to prepare and test any possible new tool needed for these kind of searches, as well as to produce any preliminary study which can assess the feasibility of these analyses using the upgraded new LHCb detector. 
%

%%%%%%%%%%%%%%%%%%%%%%%%%%%%%%%%%%%%%%%%%%%%%%%%%%%%

\section{Summary}
\label{s.summary}

We have summarised some possible trigger improvements in ATLAS, CMS, and LHCb to facilitate finding long-lived particles beyond the Standard Model at the LHC. These ideas have been discussed at LLP@LHC Community Workshops, within the experimental collaborations, and on other occasions. We encourage the use of this document to further expand the developments in Run 3 and beyond.

%%%%%%%%%%%%%%%%%%%%%%%%%%%%%%%%%%%%%%%%%%%%%%%%%%%%
\section*{Acknowledgements}

The research of David Curtin is supported in part by a Discovery Grant from the Natural Sciences and Engineering Research Council of Canada, and by the Canada Research Chair program. 
The work of Xabier Cid Vidal is supported by MINECO (Spain) through the Ramón y Cajal program RYC-2016-20073 and by XuntaGAL under the ED431F 2018/01 project. His work and that of Adrián Casais Vidal have also received financial support from XuntaGAL (Centro singular de investigación de Galicia accreditation 20192022), by European Union ERDF and ESF and by the ``María de Maeztu'' Units of Excellence program MDM-2016-0692 and the Spanish Research State Agency. The work of Adrián Casais Vidal is also supported by the Spanish Research State Agency through the program ``The Standard Model to the limits: BSM searches with LHCb'' with reference PRE2018-083399. 
Vladimir Gligorov, Christina Agapopoulou, and Lukas Calefice acknowledge support of the European Research Council under Consolidator grant RECEPT 724777.
Tova Holmes's research is supported by U.S. Department of Energy, Office of Science, Office of Basic Energy Sciences Energy Frontier Research Centers program under Award Number DE-SC0020267. 
The work of Murilo Santana Rangel was financed in part by the Coordenação de Aperfeiçoamento de Pessoal de Nível Superior – Brasil (CAPES) – Finance Code 001, CNPq and FAPERJ.
Nishita Desai acknowledges support from the Department of Science and Technology, Government of India, through grant number SB/S2/RJN-070.
Brij Jashal, Arantza Oyanguren and Louis Henry acknowledge the support from GVA and MICINN (Spain). Louis Henry acknowledges support from the ERC Consolidator Grant SELDOM G.A. 771642. 
Jose Zurita is supported by the {\it Generalitat Valenciana} (Spain) through the {\it plan GenT} program (CIDEGENT/2019/068), by the Spanish Government (Agencia Estatal de
Investigación) and ERDF funds from European Commission (Grant No. PID2020-114473GB-I00).
%%%%%%%%%%%%%%%%%
\clearpage
\bibliography{references}
\bibliographystyle{JHEP}
%%%%%%%%%%%%%%%%%

\end{document}